\documentclass[a4paper,fleqn]{cas-sc}
\usepackage[numbers]{natbib}
\usepackage{stfloats}
\usepackage{graphicx}
\usepackage{amsmath}
\usepackage{amsthm}
\usepackage{amsfonts}
\usepackage{amssymb}
\usepackage{hyperref} 
\usepackage{array}                      
\usepackage{mathtools}
\usepackage{comment}
\usepackage{xcolor}
\usepackage{mathrsfs}
\usepackage{caption}
\usepackage{subcaption}
\usepackage{tabularx}

\usepackage[font=small,labelfont=bf]{caption}
\usepackage [english]{babel}
\usepackage [autostyle, english = american]{csquotes}
\MakeOuterQuote{"}

\newcommand{\RR}{\mathbb{R}}
\newcommand{\ZZ}{\mathbb{Z}}
\newcommand{\dd}{{\tt d}}

\newcommand{\diag}[1]{\text{diag}(#1)}

\newcommand{\Span}[1]{\text{Span}(#1)}
\newcommand{\Range}[1]{\text{Range}(#1)}

\newcommand{\verts}{\mathcal{V}}
\newcommand{\edges}{\mathcal{E}}
\newcommand{\graph}{\mathcal{G}}
\newcommand{\nbd}{\mathcal{N}}
\newcommand{\adj}{\mathsf{A}}
\newcommand{\lap}{\mathsf{L}}

\newcommand{\dom}{\text{dom }}
\newcommand{\hyb}{\mathcal{H}}
\newcommand{\kyb}{\mathcal{K}}
\newcommand{\attr}{\mathcal{A}}
\newcommand{\statespc}{\mathcal{X}}

\newcommand{\distspc}{\mathcal{D}}

\newcommand{\clust}[1]{\mathcal{C}(#1)}
\newcommand{\inclust}[1]{\mathcal{I}(#1)}
\newcommand{\cluster}{\mathcal{C}}

\newcommand{\fmat}{\mathsf{F}}
\newcommand{\sproj}{\mathsf{S}}

\newcommand{\tspc}{\mathcal{T}}
\newcommand{\rspc}{\mathcal{R}}
\newcommand{\mmat}{\mathsf{M}}

\newcommand{\nbdpar}{\mathsf{N}}
\newcommand{\WV}{\mathsf{V}}
\newcommand{\WD}{\mathsf{D}}

\newcommand{\CC}{\mathsf{C}}
\newcommand{\II}{\mathsf{I}}
\newcommand{\mtt}{\mathtt{M}}
\newcommand{\ctt}{\mathtt{C}}
\newcommand{\linop}{\mathsf{T}}

\newcommand{\fred}[1]{{#1}}
\newcommand{\sean}[1]{{#1}}

\theoremstyle{definition}
\newtheorem{theorem}{Theorem}

\newtheorem{corollary}{Corollary}
             
\newtheorem{definition}{Definition}

\newtheorem{lemma}{Lemma}

\newtheorem{remark}{Remark}

\begin{document}
\let\WriteBookmarks\relax
\def\floatpagepagefraction{1}
\def\textpagefraction{.001}
\shorttitle{Consensus over Clustered Networks Using Intermittent and Asynchronous Output Feedback}
\shortauthors{F. M. Zegers and S. Phillips}

\title [mode = title]{Consensus over Clustered Networks Using Intermittent and Asynchronous Output Feedback}                      
\tnotemark[1]
\tnotetext[1]{This research is supported by the Office of Naval Research Grant N00014-21-1-2415.
Any opinions, findings, conclusions, 
or recommendations expressed in this material are those of the author(s) and do not necessarily reflect the views of the sponsoring agencies. 
Approved for public release;
distribution unlimited. Public Affairs approved AFRL-2024-3483.}
\author[1]{Federico M. Zegers}
\cormark[1]
\ead{federico.zegers@jhuapl.edu}
\affiliation[1]{organization={Johns Hopkins University Applied Physics Laboratory}, 
                city={Laurel},
                state={MD},
                postcode={20723},
                country={USA}}

\author[2]{Sean Phillips}
\affiliation[2]{organization={Air Force Research Laboratory, Space Vehicles Directorate}, 
                city={Kirtland AFB},
                state={NM}, 
                postcode={87117},
                country={USA}}

\cortext[cor1]{Corresponding author at Johns Hopkins University Applied Physics Laboratory, Laurel, MD, 20723, USA.}

\begin{abstract}
%
%In recent years, multi-agent teaming has garnered considerable interest since complex objectives, such as intelligence, surveillance, and reconnaissance, can be divided into multiple cluster-level sub-tasks and assigned to a cluster of agents with the appropriate functionality.
%
%Yet, coordination and information dissemination between clusters may be necessary to accomplish a desired objective.
%
Distributed consensus protocols provide a mechanism for spreading information within clustered networks, allowing agents and clusters to make decisions without requiring direct access to the state of the ensemble.
\fred{In this work}, we propose a strategy for achieving system-wide consensus in the states of identical linear time-invariant systems coupled by an undirected graph whose directed sub-graphs are available only at sporadic times. 
Within this work, the agents of the network are organized into pairwise disjoint clusters, which induce sub-graphs of the undirected parent graph. 
Some cluster sub-graph pairs are linked by an inter-cluster sub-graph, where the union of all cluster and inter-cluster sub-graphs yields the undirected parent graph.
Each agent utilizes a distributed consensus protocol with components that are updated intermittently and asynchronously with respect to other agents \fred{and inter-clusters}.
The closed-loop ensemble dynamics is modeled as a hybrid system, and a Lyapunov-based stability analysis yields sufficient conditions for rendering the agreement subspace (consensus set) globally exponentially stable.
Furthermore, an input-to-state stability argument demonstrates the consensus set is robust to a \fred{large} class of perturbations. 
A numerical simulation considering both nominal and perturbed scenarios is provided for validation purposes. 
\end{abstract}

\begin{keywords}
Multi-Agent Systems \sep 
Decentralized Control \sep 
Networked Systems \sep
Stability of Hybrid Systems \sep
Lyapunov Methods
\end{keywords}

\maketitle

\section{Introduction}
\subsection{Motivation}
The Internet of Things (IoT) and advancements in data handling within network systems have created a paradigm shift in how the internet and distributed platforms are used today~\cite{Li.Hu.ea2021}.
Consumer electronics and household devices such as cameras, watches, televisions, and refrigerators are now equipped with WiFi or LAN functionality\fred{,} granting the user access to nearly instantaneous data and wider control of their environment~\cite{Chang.Srirama.ea2019}. 
Connected devices, like those listed above, possess a natural graphical structure called a cluster-based network~\cite{Wang.Xiao2006} or small-world network~\cite{Zaidi2013,Telesford.Joyce.ea2011,Watts.Strogatz1998}.
Although sophisticated, the technology enabling current small-world networks will need to evolve so that it can address imminent challenges.
For instance, as more devices gain internet connectivity and transmit more data due to improved capability, the resulting small-world networks will increase in complexity and require enhanced data handling and control protocols to meet demands.
Bottlenecks introduced by finite network resources, e.g., bandwidth and throughput, and the need to perform control using limited information are ever-present constraints for multi-agent systems (MASs), especially those \fred{possessing} dynamics, such as vehicles.

Given their wide applicability and utility in coordination~\cite{Mesbahi.Egerstedt2010}, we are interested in distributed consensus protocols for MASs.
Specifically, we focus on consensus over clustered networks while utilizing intermittent and asynchronous output feedback.
Data-handling protocols that embrace \fred{intermittency, asynchrony, and limited information are better suited for large-scale networks than policies requiring periodically and synchronously derived full state feedback}.
In practice, certain components of cyber-physical systems may operate at a constant rate, for example, electromechanical sensors and digital processors.
\fred{Hence}, some information needed for control may be available periodically, motivating the development of controllers that \fred{can} seamlessly integrate with such components.
\fred{Intermittency can also promote the efficient use of limited computation and communication resources.}

\subsection{Background}
The consensus problem for MASs, in both the continuous-time and discrete-time settings, has been well-studied in~\cite{OlfatiSaber.Fax.ea2007}. 
Furthermore, the survey papers in~\cite{Qin.Ma.ea2017a, Li.Tan2019, Dorri.Kanhere.ea2018} provide additional information on MASs and consensus. 
Recently, however, there has been significant research on consensus where the communication between agents and control updates within agents are periodic, intermittent, and/or event-triggered~\cite{Ding.Han.ea2017,Nowzari.Garcia.ea2019,Ge.Han2016, Phillips.Sanfelice2024, Zegers.Guralnik.ea2022a, Phillips.Sanfelice2019a,Zegers.Guralnik.ea2024,Fan.Yang.ea2016}. 
These works were motivated by the efficient use of limited resources and the implementation challenges inherent to cyber-physical systems caused by, for example, sensor sampling rates, radio broadcast rates, and actuator constraints.
Yet, there is a lack of research focusing on consensus over clustered networks, which is somewhat surprising since it is natural to solve complicated problems with cooperative teams of specialized agents.
In fact, clustering in network systems is quite common in engineering applications~\cite{Wirasanti.Ortjohann.ea2012,Cheng.Scherpen2018,SanchezGarcia.Fennelly.ea2014,Wang.Xiao2006,Bandyopadhyay.Coyle2004,Jayawardene.Herath.ea2020}. 

The results in~\cite{Qin.Yu2013,Ge.Han.ea2018,Pham.Messai.ea2020a,Pham.Messai.ea2020b,Luo.Ye2022,Zegers.Phillips.ea2021,Nino.Zegers.ea2023} delve into the control of MASs over clustered networks.
In~\cite{Qin.Yu2013}, the authors develop a distributed controller enabling $p$-cluster consensus for a MAS of identical linear time-invariant (LTI) systems coordinating over a static or switching graph.
Given a graph encoding a MAS whose vertex set is partitioned into $p$ elements (i.e., clusters), $p$-cluster consensus is said to occur whenever the agents in each cluster achieve consensus, and the consensus value between clusters is distinct.
Note that $p$-cluster consensus is accomplished using continuous state feedback.
In~\cite{Ge.Han.ea2018}, the authors investigate the formation control problem for clustered networks of identical LTI systems.
The authors develop a distributed controller that enables $M$ clusters to assemble distinct rigid formations using intermittent state feedback while compensating for communication delays.

The result in~\cite{Pham.Messai.ea2020a} presents a distributed consensus controller for a clustered network of identical LTI systems, where each agent leverages a mixture of continuous and intermittent output feedback.
Each cluster is assigned a single leader, and, within each cluster, all agents may continuously communicate output information with their neighbors.
However, only the leaders of each cluster may communicate intermittently with neighboring leaders, and this is the means by which information is spread across clusters.
In addition, all leaders communicate with their neighboring leaders at the same isolated times.
A Luenberger-like observer produces state estimates from output information, which are then used for control. 
Albeit not written, the analysis in~\cite{Pham.Messai.ea2020a} renders the agreement subspace for the entire ensemble globally exponentially stable \fred{(GES)}.
Leveraging analogous arguments and cluster structures found in~\cite{Pham.Messai.ea2020a}, the result in~\cite{Pham.Messai.ea2020b} develops a distributed controller enabling a clustered network of identical LTI systems to assemble a single rigid formation using a mixture of continuous and intermittent state feedback. 
In~\cite{Luo.Ye2022}, the authors build on the result of~\cite{Qin.Yu2013} and identify the general graph property that supports $p$-cluster consensus independent of coupling strength.

\subsection{Comparison to Previous Work}
In~\cite{Zegers.Phillips.ea2021}, we build on~\cite{Pham.Messai.ea2020a} and develop a distributed controller for a clustered network of identical LTI systems that enables consensus.
Like~\cite{Pham.Messai.ea2020a}, communication within clusters is continuous \fred{while} communication between clusters is intermittent.
\fred{Yet}, unlike~\cite{Pham.Messai.ea2020a}, \fred{the result in~\cite{Zegers.Phillips.ea2021} does not employ a leader-follower strategy} and more general partitions of the communication graph are considered.
As a result, more than one agent in a cluster can have a neighbor in a different cluster.
\fred{Two} pairs of distinct clusters may exchange information at different rates (i.e., asynchronously) \fred{in~\cite{Zegers.Phillips.ea2021}, which is not supported in~\cite{Pham.Messai.ea2020a}}.
Still, it is important to \fred{note} that~\cite{Pham.Messai.ea2020a} considers output feedback while~\cite{Zegers.Phillips.ea2021} only considers state feedback.

Our result in~\cite{Nino.Zegers.ea2023} is an extension of~\cite{Zegers.Phillips.ea2021}\fred{, which} develops a distributed controller for a clustered network of identical LTI systems \fred{facilitating} consensus using \fred{only} output feedback.
Similar to the results in~\cite{Pham.Messai.ea2020a} and~\cite{Zegers.Phillips.ea2021}, 
communication \fred{within} clusters is continuous while communication between clusters is intermittent.
Rather than broadcast state information between neighbors as in~\cite{Zegers.Phillips.ea2021}, \fred{the result in~\cite{Nino.Zegers.ea2023} broadcasts state estimates computed using a Luenberger-like observer instead.}
\fred{This observer leverages intermittently sampled outputs, model knowledge, and control input measurements to compute state estimates.}

Unfortunately, the stability analysis in~\cite{Nino.Zegers.ea2023} suffers from an artificial defect.
To demonstrate the desired set $\attr$ is \fred{GES}, the stability analysis required all Luenberger-like observers to undergo synchronous jumps.
The price to pay for this mathematical convenience is that all agents must synchronously sample their output, which is \fred{impractical} for large-scale networks.

\subsection{\fred{Control Strategy Description \& Contributions}}
Motivated by our previous work in~\cite{Zegers.Phillips.ea2021,Nino.Zegers.ea2023}, we introduce a \fred{new} distributed consensus controller for a clustered network of identical LTI systems \fred{that leverages} intermittent and asynchronous output feedback \fred{both within and between clusters}.
The clustered network is modeled as a connected, undirected, and static graph, \fred{which is decomposed into cluster and inter-cluster sub-graphs}.\footnote{\fred{Such a structure can be used to encode a team of teams, where clusters can describe teams of agents and inter-clusters can describe connections between these teams.}} 
Each agent and inter-cluster \fred{sub-graph} is assigned a \fred{unique} timer mechanism that dictates when measurement and control update events \fred{occur}.
Each timer mechanism \fred{is responsible for generating} isolated events times, \fred{where the difference between consecutive event times can be bounded above and below by user-defined parameters, allowing for the \textit{a priori} exclusion of Zeno behavior, the satisfaction of hardware constraints (e.g., sensor sampling rate), and the satisfaction of the maximum allowable transmission interval (MATI)~\cite{Phillips.Sanfelice2019}.}
Within each cluster, each agent intermittently measures the outputs of their same-cluster neighbors and themselves when instructed by their timer mechanism.
Similarly, within an inter-cluster, each agent intermittently measures the outputs of their same-inter-cluster neighbors and themselves when instructed by the inter-cluster timer mechanism.
Thus, the measurement of output information is completely intermittent throughout the entire network, unlike~\cite{Pham.Messai.ea2020a,Zegers.Phillips.ea2021,Nino.Zegers.ea2023}.
Since these timer \fred{mechanisms} are independent, the flow of output information within clusters and \fred{inter-clusters may be asynchronous, resulting in the directed flow of output information along the edges of an undirected graph.}
\fred{Every agent computes a consensus-like estimate for their assigned cluster and each inter-cluster the agent is a member of, which evolve in an open-loop fashion during flows and are impulsively reset with the appropriate output measurements during jumps as determined by the corresponding timer mechanism.
Each agent uses their consensus-like estimates to compute a control input at each moment in time, which is shown to enable consensus while only using intermittently and asynchronously available output measurements.
Such estimators provide a simpler alternative to the Luenberger-like observes utilized in~\cite{Nino.Zegers.ea2023}, supplanting model-based state reconstruction with model-less output-based estimation.}

When considering consensus or synchronization problems \fred{over finite-dimensional metric spaces}, it is customary to examine disagreement dynamics, where disagreement is usually defined \fred{through} a projection onto the \fred{orthogonal complement of the} agreement subspace.
However, such a projection is not full rank and can prevent the satisfaction of sufficient conditions expressed as matrix inequalities ensuring the stability of the agreement subspace.
To overcome this challenge, \fred{we develop an alternative disagreement metric based on the image of a full-rank operator.
This alternative disagreement metric is used to derive a closed-loop disagreement dynamics, consisting of continuous-time and discrete-time behaviors, modeled as a hybrid system of the form in~\cite{Goebel.Sanfelice.ea2012}.
The consensus objective is cast as a set stabilization problem, where the proposed hybrid system is shown to be nominally well-posed, the desired set is shown to be compact and GES for the proposed hybrid system via a Lyapunov-based stability analysis, and the desired set is shown to be robust to Lebesgue measurable and locally essentially bounded perturbations through an input-to-state stability argument.\footnote{\fred{The nominal well-posedness of the proposed hybrid system, compactness of the desired set, and the GES of the desired set imply this set is robust to vanishing perturbations}.}
Such results are not considered in~\cite{Pham.Messai.ea2020a,Zegers.Phillips.ea2021,Nino.Zegers.ea2023} nor is the GES of a set encoding the consensus objective achieved by a control strategy leveraging only intermittent and asynchronous information.
To the best of our knowledge, the contributions of this work to the literature are the following:
\begin{enumerate}
    \item A distributed consensus controller for clustered networks of identical LTI systems utilizing only intermittent and asynchronous output feedback;
    \item Stemming from Item 1, a nominally well-posed hybrid system with complete non-Zeno solutions and a GES compact attractor;
    \item Certificates of robustness to vanishing perturbations via standard results in~\cite{Goebel.Sanfelice.ea2012} and Lebesgue measurable and locally essentially bounded perturbations via an input-to-state stability proof.
\end{enumerate}
}
The theoretical development is validated in simulation, where we provide an online path planner for a rendezvous application.

\subsection{Organization}
Section~\ref{sec: Preliminaries} provides information on notation, graphs, cluster and inter-cluster sub-graphs, and hybrid systems. 
Section~\ref{sec: Problem Statement} describes the specific problem we are aiming to solve.
The proposed solution and closed-loop hybrid \fred{system} are presented in Section~\ref{sec: Modeling}. 
In Section~\ref{sec: Stability Analysis}, we provide the main results of the paper in the form of sufficient conditions for the nominal system and, in Section~\ref{sec: Robustness}, we extend these results to consider \fred{a class} of perturbations. 
Lastly, we illustrate the results through numerical simulations in Section~\ref{sec: Numerical}.

\section{Preliminaries}\label{sec: Preliminaries}
\subsection{Notation} \label{sec: Notation}
Let $\RR$ and $\ZZ$ denote the set of reals and integers, respectively. 
For a constant $a\in\RR$, let $\RR_{\geq a}\coloneqq[a,\infty)$,
$\RR_{>a}\coloneqq(a,\infty)$, $\ZZ_{\geq a}\coloneqq\RR_{\geq a}\cap\ZZ$, and $\ZZ_{>a}\coloneqq\RR_{>a}\cap\ZZ$.
For $p,q\in\ZZ_{>0}$, the $p\times q$
zero matrix and the $p\times 1$ zero column vector are respectively denoted by
$0_{p\times q}$ and $0_p$.
The $p\times p$ identity
matrix and the $p\times 1$ column vector with all entries being one are denoted by $I_p$ and $1_p$, respectively. 
The Euclidean norm of $r\in\RR^p$
is denoted by $\Vert r\Vert \coloneqq\sqrt{r^\top r}$.
For $x,y\in\RR^p$, the inner product between $x$ and $y$
is denoted by $\langle x,y\rangle\coloneqq x^\top y$. 
For $M\in\ZZ_{\geq 2}$, let $[M]\coloneqq\{1,2,...,M\}$.
The Kronecker product between $A\in\RR^{p\times q}$ and $B\in\RR^{u\times v}$, where $u,v\in\ZZ_{>0}$, is denoted by $A\otimes B\in\RR^{pu\times qv}$.
The maximum and minimum eigenvalues of a real symmetric matrix $A\in\RR^{n\times n}$ are denoted by $\lambda_{\max}(A)$ and $\lambda_{\min}(A)$, respectively.
The block diagonal matrix with general blocks $G_1,G_2,...,G_p$ is denoted by $\text{diag}(G_1,G_2,...,G_p)$.
The distance of a point $r\in\RR^p$ to the set $S\subset\RR^p$ is given by $\vert r\vert_{S}\coloneqq\inf\{\Vert r-s\Vert\colon s\in S\}\in\RR_{\geq 0}$\fred{, and let} $r+S\coloneqq \{r+s\in\RR^p\colon s\in S\}$.
For a collection of vectors $\{z_1,z_2,...,z_p\}\subset\RR^q$, let $(z_k)_{k\in[p]}\coloneqq [z_1^\top,z_2^\top,...,z_p^\top]^\top\in\RR^{pq}$.
Similarly, for $x\in\RR^p$ and $y\in\RR^q$, let $(x,y)\coloneqq [x^\top,y^\top]^\top\in\RR^{p+q}$.
For any sets $A$ and $B$, a function $f$ of $A$ with values in $B$ is denoted by $f\colon A\to B$, whereas $f\colon A\rightrightarrows B$ refers to a set-valued function $f\colon A\to 2^B$.
The $k^\text{th}$ standard basis vector of $\RR^N$, for $N\in\ZZ_{>0}$, is denoted by $\mathtt{e}_k$.
Similarly, let $\tilde{\mathtt{e}}_k$ denote the $k^\text{th}$ standard basis vector of $\RR^{N-1}$.
The range (i.e., image) of a matrix $A\in\RR^{N\times N}$ is denoted by $\Range{A} \coloneqq \{y\in\RR^N\colon y = A x \text{ for some } x\in\RR^N\}$.
For any $x\in\RR$, let $\lfloor x\rfloor\coloneqq\sup\{c\in\ZZ\colon c\leq x\}$.
\fred{
Given a symmetric real-valued matrix $A$, the notation $A>\mathbf{0}$, $A\geq \mathbf{0}$, $A<\mathbf{0}$, and $A\leq \mathbf{0}$ imply $A$ is positive definite, positive semi definite, negative definite, and negative semi definite, respectively.}

\subsection{Graphs}
Let $\graph\coloneqq(\verts,\edges)$ be a graph on $N\in\ZZ_{\geq2}$ nodes, where $\verts\coloneqq[N]$ denotes the node set and $\edges\subseteq\verts\times\verts$ denotes the edge set.
If $(p,q)\in\edges$ implies $(q,p)\in\edges$ for all distinct nodes $p,q\in\verts$, then the graph $\graph$ is said to be undirected. 
A path exists between nodes $p,q\in\verts$ if there is a sequence of distinct nodes such that $(v_{0}=p,...,v_{k}=q)$ for $k\in\ZZ_{\geq0}$, $(v_{s-1},v_{s})\in\edges$, and $s\in[k]$.
The graph $\graph$ is said to be connected if there is a path joining any two distinct nodes in $\verts$.
The neighbor set of node $p$ is denoted by $\nbd_p\coloneqq\{q\in\verts\setminus\{p\}:(p,q)\in\edges\}$.
The adjacency matrix of $\graph$ is denoted by $\adj\coloneqq[a_{pq}]\in\RR^{N\times N}$, where $a_{pq}=1$ if and only if $(p,q)\in\edges$, and $a_{pq}=0$ otherwise.
Self-edges are not employed in this work, that is, $a_{pp}\coloneqq 0$ for all $p\in\verts$.
The degree matrix of $\graph$ is denoted by $\Delta\coloneqq\diag{\adj\cdot 1_N}\in\RR^{N\times N}$.
The Laplacian matrix of $\graph$ is denoted by $\lap\coloneqq\Delta-\adj\in\RR^{N\times N}$.
The following result enables the stability analysis provided in Section~\ref{sec: Stability Analysis}.
\begin{lemma} \label{lemma: L and S Identities}
    If $\graph$ is static, undirected, and connected, then there exists an orthonormal basis $\beta\coloneqq\{v_1,v_2,...,v_N\}\subset\RR^N$ for $\Range{\lap}$ such that $v_1 = (\sqrt{N}/N) 1_N$.
    Consider the matrix $\WV\coloneqq [v_2,v_3,...,v_N]\in\RR^{N\times N-1}$ and projection $\sproj\coloneqq I_N - 1_N 1_N^\top / N \in\RR^{N\times N}$.
    Then,
    \begin{align}
        \lap &= \WV\WD\WV^\top, \ \WV^\top\WV = I_{N-1}, \label{eqn: Laplacian Identity} \text{ and }\\
        \sproj &= \WV\WV^\top \label{eqn: S Identity}
    \end{align}
    for some diagonal, positive definite $\WD\in\RR^{N-1\times N-1}$.
    \hfill$\triangle$
\end{lemma}
The matrix $\sproj$ is a projection whose image is the orthogonal complement of the agreement subspace $\Span{1_N}$. 
That is, for any vector $z\in\RR^N$, $z$ may be decomposed into the sum of two orthogonal vectors, i.e., $z = z^\Vert + z^\perp$, where $z^\Vert \coloneqq (1_N 1_N^\top / N)z$ and $z^\perp \coloneqq \sproj z$.
In addition, for any symmetric Laplacian matrix $\mathcal{L}$, one can show by direct computation that $\sproj\mathcal{L}=\mathcal{L}=\mathcal{L}\sproj$.
\begin{proof}[Proof of Lemma~\ref{lemma: L and S Identities}]
    See Appendix~\ref{appen: L and S Identities}.
\end{proof}

\subsection{Cluster and Inter-Cluster Sub-Graphs} \label{sec: Cluster and Inter-Cluster Sub-Graphs}
\fred{In this section, we introduce the precise notions of cluster and inter-cluster sub-graphs leading to the clustered networks discussed in this work.
Please see~\cite{Pham.Messai.ea2020a,Pham.Messai.ea2020b,Shi.Hu.ea2023} for additional information on clustered networks.}
Let $\graph=(\verts,\edges)$ be a static, undirected, and connected graph.
Let $\clust{\verts}\coloneqq\{\verts_1,\verts_2,...,\verts_M\}$ be a fixed partition of the set $\verts$, where $M\in [N]$ is some constant.
Consequently, $\verts_p\subseteq\verts$ for each $p\in[M]$, $\verts_p\cap\verts_q=\varnothing$
for each distinct pair $p,q\in[M]$, and $\cup_{p\in[M]}\verts_p = \verts$.
The sets $\clust{\verts}$ and $\verts_q$, for any $q\in\verts$, are referred to as the \textit{cluster set of $\graph$} and \textit{cluster} $q$, respectively.
Since $\clust{\verts}$ is an arbitrary partition of $\verts$, for every $p\in [M]$, cluster $p$ contains at least one node, i.e., $\vert \verts_p \vert \geq 1$.
Each cluster induces a sub-graph of $\graph$, i.e., for each $p\in[M]$, the induced sub-graph of cluster $p$ is given by $\graph[\verts_p]\coloneqq(\verts_p,\edges[\verts_p])$,
such that $\edges[\verts_p]\coloneqq\{(r,s)\in\edges\colon r,s\in\verts_p\}$.
Since $\graph$ is undirected, $\graph[\verts_p]$
is undirected.
In this work, we only consider cluster sets $\clust{\verts}$ where, for each $p\in [M]$, $\graph[\verts_p]$ is connected.
Moreover, since $\clust{\verts}$ is a fixed partition of $\verts$, for each $p\in\verts$, node $p$ is assigned to one cluster for all time.
In particular, for each $p\in\verts$, let $\cluster_p$ denote the cluster to which node $p$ belongs, i.e., $p\in\cluster_p$ and $\cluster_p=\verts_q$ for some $\verts_q\in\clust{\verts}$.
The set of neighbors of node $p$ within $\cluster_p$ is, therefore, given by $\nbd_p\cap\cluster_p$.

\begin{figure}[t]
\centering 
\includegraphics[width=0.49\columnwidth]{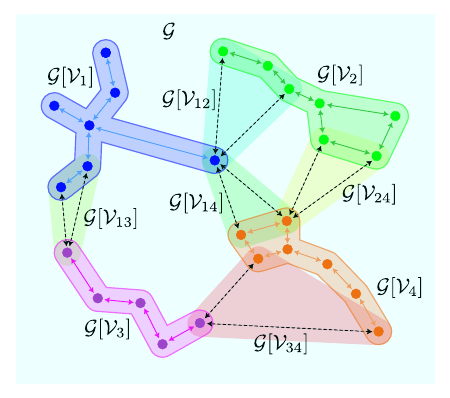} 
\caption{Within the undirected graph $\graph$, as denoted by the disks and double-headed arrows, there are four sub-graphs induced by clusters and five sub-graphs induced by inter-cluster.
Note that solid arrows represent edges within clusters, while dashed arrows represent edges between clusters, i.e., within inter-clusters.
The cluster sub-graphs of $\graph$ are $\graph[\verts_1]$, $\graph[\verts_2]$, $\graph[\verts_3]$, and $\graph[\verts_4]$.
The inter-cluster sub-graphs of $\graph$ are $\graph[\verts_{12}]$, $\graph[\verts_{13}]$, $\graph[\verts_{14}]$, $\graph[\verts_{24}]$, and $\graph[\verts_{34}]$
By construction, the union of all sub-graphs returns the original graph $\graph$.
}
\label{fig: Clustered graph}
\end{figure}

For any distinct pair $p,q\in[M]$, let $\verts_{pq} \coloneqq \verts(p,q)\cup \verts(q,p)$, such that $\verts(p,q)\coloneqq \{ r\in\verts_p \colon \exists_{s\in\verts_q} \ (r,s)\in\edges\}$.
If $\verts_{pq}\neq\varnothing$, then the set $\verts_{pq}$ is referred to as \textit{inter-cluster} $pq$.
To simplify the development, we only consider the case where there is at most one inter-cluster between each pair of distinct clusters.
The \textit{inter-cluster set of $\graph$}, as determined by $\clust{\verts}$, is denoted by $\inclust{\verts}\coloneqq \{\verts_{pq}\subset\verts\colon \forall_{p\neq q\in\verts} \ \verts_{pq}\neq\varnothing \}$.
Since $\vert\clust{\verts}\vert=M$ and there can be at most one inter-cluster between each distinct pair of clusters, $\vert \inclust{\verts}\vert = M^\ast\in \{0,1,...,M(M-1)/2\}$.
For each distinct pair of $p,q\in[M]$, the sub-graph of $\graph$ induced by inter-cluster $pq$ is denoted by $\graph[\verts_{pq}]\coloneqq(\verts_{pq},\edges_{pq})$, where $\edges_{pq}\coloneqq\edges[\verts_{pq}]\setminus(\edges[\verts_p]\cup\edges[\verts_q])$.
Further, $\graph[\verts_{pq}]$ is undirected by construction.
Figure~\ref{fig: Clustered graph} illustrates an example of a graph $\graph$ that is divided into cluster and inter-cluster sub-graphs.
Since $\clust{\verts}$ and $\graph$ are fixed, one has that $\inclust{\verts}$ is fixed as well.
Thus, if node $p\in\verts$ belongs to an inter-cluster, then it does so for all time.
Moreover, it is possible for a node to belong to multiple inter-clusters simultaneously.
The set of neighbors of node $p$ in inter-cluster $rs$ is, therefore, given by $\nbd_p\cap\verts_{rs}$.

To facilitate the analysis, $N\times N$ augmented adjacency and Laplacian matrices for a general sub-graph of $\graph$ are introduced next.
For any non-empty set $S\subseteq\verts$, let $\adj[S]\coloneqq[\mathsf{a}_{pq}]\in\RR^{N\times N}$ with
\begin{equation*}
    \mathsf{a}_{pq} \coloneqq \Bigg\{
        \begin{aligned}
            &1, & (p,q) \in \edges[S]\\
            &0, & \text{otherwise}
        \end{aligned}
\end{equation*}
denote the augmented adjacency matrix of $\graph[S]$.
Therefore, the augmented adjacency matrix of cluster $p$ and inter-cluster $pq$ are $\adj[\verts_p]$ and $\adj[\verts_{pq}]$, respectively.
Each augmented adjacency matrix has a corresponding augmented Laplacian matrix.
The augmented Laplacian matrix of $S\subseteq\verts$ is $\lap[S]\coloneqq \diag{\adj[S]\cdot 1_N} - \adj[S]\in\RR^{N\times N}$.
Consequently, the augmented Laplacian matrix of cluster $p$ and inter-cluster $pq$ are $\lap[\verts_p]$ and $\lap[\verts_{pq}]$, respectively.

For convenience, we introduce and henceforth use an alternative labeling scheme for the inter-clusters of $\graph$ as determined by a cluster set $\clust{\verts}$.
Let $\mathtt{b}\colon[M]\times[M]\to[M^\ast]$ be a bijection.
Given a cluster set $\clust{\verts}$ with cardinality $M$, there are exactly $M^\ast$ pairs of distinct clusters with an inter-cluster.
Hence, for some $p,q\in[M]$ with $p\neq q$, there exists a unique $r\in[M^\ast]$ such that $\mathtt{b}(p,q)=r$.
By inductively using the bijection $\mathtt{b}$, let $\verts^r\coloneqq \verts_{pq}$ for all $p,q\in[M]$ with $p\neq q$ such that $\mathtt{b}(p,q)=r$.
Thus, the sub-graph and augmented Laplacian matrix of inter-cluster $pq$, for example, can also be written as $\graph[\verts^r]$ and $\lap[\verts^r]$, respectively.
A static, undirected, and connected graph $\graph$ with cluster and inter-cluster sub-graphs determined by cluster set $\clust{\verts}$ and inter-cluster set $\inclust{\verts}$, respectively, is henceforth referred to as \textit{clustered network}.

For each $p\in\verts$, the neighbor set of agent $p$, that is, $\nbd_p$, can be partitioned using the cluster set $\clust{\verts}$ and inter-cluster set $\inclust{\verts}$.
The partition of $\nbd_p$ is given by $\nbdpar_p\coloneqq\{\nbd_p\cap\cluster_p, \ \nbd_p\cap\verts^1, \ \nbd_p\cap\verts^2, ..., \nbd_p\cap\verts^{M^\ast}\}$, where $S_1\cap S_2=\varnothing$ for any distinct $S_1,S_2\in\nbdpar_p$ and 
\begin{equation} \label{eqn: neighbor set identity}
    \nbd_p = (\nbd_p\cap\cluster_p)\cup\big(\cup_{r\in[M^\ast]}(\nbd_p\cap\verts^r)\big).
\end{equation}
Note, for any connected graph $\graph$, it may be possible for $\nbd_p\cap\cluster_p = \varnothing$ for some agent $p\in\verts$ and/or $\nbd_p\cap\verts^r=\varnothing$ for some inter-cluster $r\in [M^\ast]$.

\subsection{Hybrid Systems}

A hybrid system $\hyb$ with data $(C,f,D,G)$ is defined as
\begin{equation*}
    \hyb \colon    
    \Bigg\{
    \begin{aligned}
        \dot{\xi} &= f(\xi), \ & \xi\in C\phantom{,} \\
        \xi^+ &\in G(\xi), \ & \xi\in D,
    \end{aligned}
\end{equation*}
where $f:\RR^n\to\RR^n$ denotes the flow map, $C\subset\RR^n$ denotes the flow set, $G:\RR^n\rightrightarrows\RR^n$ denotes the jump map, and $D\subset\RR^n$ denotes the jump set.\footnote{\fred{Within the context of multi-agent systems, the hybrid system $\hyb$ will be used to model the continuous-time and discrete-time behaviors of the ensemble.
In particular, the flow set $C$ and jump set $D$ will be used to determine when the state of the ensemble, namely, $\xi$, may flow according to the flow map $f$ and/or jump according to the jump map $G$, respectively.
For example, if the state of the ensemble is in the jump set, i.e., $\xi\in D$, then the state variable of the ensemble may jump according to $\xi^+\in G(\xi)$, where some variables in $\xi$ may change discretely while other variables in $\xi$ may be mapped to their current value under an identity transformation (since they may only change in continuous time). 
}}
A set $\Omega\subset\RR_{\geq 0}\times\ZZ_{\geq 0}$ is a hybrid time domain, if there is a non-decreasing sequence of non-negative reals $(t_j)_{j=0}^m$ with $m\in\ZZ_{\geq 0}\cup\{\infty\}$, $t_0=0$, and $t_m\in\RR_{\geq 0}\cup\{\infty\}$, such that $\Omega=\cup_{j=1}^m(I_j\times\{j-1\})$, where all the intervals $I_j$, for $j<m$, are of the form $[t_{j-1},t_j]$, and $I_m$ is of the form $[t_{m-1},t_m]$ or $[t_{m-1},t_m)$ when $m<\infty$.
Note that $t_m=\infty$ is allowed when $m<\infty$; for $m=\infty$ there is no $t_m$.
The time $t_j$ indicates the $j^{\text{th}}$ instant the state $\xi$ jumps, where the value of $\xi$ after a jump is denoted by $\xi^+$.
A  hybrid arc $\phi$ is a function $\phi:\dom\phi\to\RR^n$, such that $\dom\phi\subset\RR_{\geq 0}\times\ZZ_{\geq 0}$ is a hybrid time domain with jump sequence $(t_j)_{j=0}^m$, and $\phi$ is a locally absolutely continuous function on $I_j$, for every $j$.
A solution of $\hyb$ is a hybrid arc $\phi$ such that, for all $j>0$, $\phi(t,j-1)\in C$ and $\dot{\phi}(t,j-1)=f(\phi(t,j-1))$ for almost all $t\in I_j$; and $\phi(t_{j-1},j-1)\in D$ and $\phi(t_{j-1},j)\in G(\phi(t_{j-1},j-1))$.
A solution $\phi$ to $\hyb$ is called maximal if $\phi$ cannot be extended, i.e., there does not exist a distinct solution $\psi$ to $\hyb$ such that $\dom\phi\subseteq\dom\psi$ and $\phi(t,j)=\psi(t,j)$ for all $(t,j)\in\dom\phi$.
A solution $\phi$ is called complete if $\dom\phi$ is unbounded.
A hybrid system $\hyb$ with data $(C,f,D,G)$ is said to satisfy the hybrid basic conditions if it satisfies~\cite[Assumption 6.5]{Goebel.Sanfelice.ea2012}.
The following definition originates from~\cite[Section II]{Teel.Forni.ea2013}.
\begin{definition}  \label{def: GES} 
    Let $\hyb$ be a hybrid system on $\statespc\subset\RR^n$ with data $(C,f,D,G)$.
    A closed set $\attr\subset\statespc$ is said to be \textit{globally exponentially stable} for the hybrid system $\hyb$ if there exist constants $\kappa, \alpha\in\RR_{>0}$ such that every maximal solution $\phi$ of $\hyb$ is complete and satisfies
    \begin{equation*}
        \vert\phi(t,j)\vert_\attr \leq \kappa \exp({-\alpha (t+j)}) \vert\phi(0,0)\vert_\attr   
    \end{equation*}
\noindent for all $(t,j) \in\dom\phi$.
\hfill$\triangle$
\end{definition}
When considering perturbations on $\hyb$, input-to-state stability \fred{offers a convenient tool for analysis.}
The following definition \fred{is a slight modification of}~\cite[Definition 2.2]{Li.Phillips.ea2018}.
\begin{definition}  \label{def: ISS} 
    The hybrid system with state $z$ and \fred{perturbation} $\fred{d}$ given by
    \begin{equation} \label{eqn: ISS hybrid system}
    \begin{aligned}
        \dot{z} &= f(z,\fred{d}), \ & (z,\fred{d})\in C \\
        z^+ &\in G(z,\fred{d}), \ & (z,\fred{d})\in D
    \end{aligned}
    \end{equation}
    is said to be \fred{input-to-state stable (ISS)} with respect to a compact set $\attr$ if there exist $\beta\in\mathcal{KL}$\footnote{A continuous function $\beta\colon\RR_{\geq 0}\times\RR_{\geq 0}\to\RR_{\geq 0}$ belongs to class $\mathcal{KL}$ if 1) $\beta(r,s)$ is non-decreasing in $r$, 2) $\lim_{r\to 0^+}\beta(r,s)=0$ for each fixed $s\in\RR_{\geq 0}$, 3) $\beta(r,s)$ is non-increasing in $s$, and 4) $\lim_{s\to\infty}\beta(r,s)=0$ for each fixed $r\in\RR_{\geq 0}$.} and $\gamma\in\mathcal{K}$\footnote{A continuous function $\gamma\colon[0,a]\to\RR_{\geq 0}$ with $a>0$ belongs to class $\mathcal{K}$ if 1) $\gamma$ is strictly increasing and 2) $\gamma(0)=0$.} such that, for each $\phi(0,0)\in\RR^n$, every solution pair $(\phi,\fred{d})$ satisfies
    \begin{equation*}
        \vert\phi(t,j)\vert_\attr \leq \max\{\beta(\vert\phi(0,0)\vert_\attr, t+j), \gamma(\vert \fred{d}\vert_{\infty}) \}  
    \end{equation*}
\noindent for all $(t,j) \in\fred{\dom(\phi,d)}$.
\hfill$\triangle$
\end{definition}
\fred{Note, $(\phi,\fred{d})$ is said to be a solution pair} of~\eqref{eqn: ISS hybrid system} if, for each $j\in\ZZ_{\geq 0}$, the \fred{mapping} $t\mapsto\phi(t,j)$ is locally absolutely continuous, $t\mapsto \fred{d}(t,j)$ is Lebesgue measurable and locally essentially bounded over $\{t\colon (t,j)\in\text{dom}(\phi,\fred{d})\}$, $\dom\phi=\dom \fred{d}$ where $\text{dom}(\phi,\fred{d})\coloneqq\dom\phi $ represents the hybrid time domain of $(\phi,\fred{d})$, $(\phi(t,j-1),\fred{d}(t,j-1))\in C$ and $\dot{\phi}(t,j-1) = f(\phi(t,j-1),\fred{d}(t,j-1))$ for almost all $t\in I_j$, and $(\phi(t_{j-1},j-1),\fred{d}(t_{j-1},j-1))\in D$ and $\phi(t_{j-1},j) \in G(\phi(t_{j-1},j-1),\fred{d}(t_{j-1},j-1))$. 
\fred{Note, $\vert\cdot\vert_{\infty}$ is defined in~\cite[Right Column of Page 48]{Cai.Teel2009}.
The class of perturbations defined within Definition~\ref{def: ISS} will serve as the non-vanishing perturbations discussed in Section~\ref{sec: Robustness}.}

\section{Problem Formulation} \label{sec: Problem Statement}
\fred{Consider a MAS of $N\in\ZZ_{\geq 2}$ agents indexed by the elements of $\verts$, where the dynamics of agent $p\in\verts$ is 
\begin{equation} \label{eqn: Agent p dynamics}
    \dot{x}_p = Ax_p + Bu_p,\quad y_p = Hx_p.
\end{equation}
Note, $x_p\in\RR^n$, $u_p\in\RR^d$, and $y_p\in\RR^m$ represent the state variable, control input, and output of the agent, respectively.
Furthermore, $A\in\RR^{n\times n}$, $B\in\RR^{n\times d}$, and $H\in\RR^{m\times n}$ represent the known state, control effectiveness, and output matrices, respectively.
To facilitate cooperation, the agents may exchange information via a communication network modeled by a static, undirected, and connected graph $\graph=(\verts,\edges)$.
Depending on the application, a MAS may need to be divided into several smaller teams, which may require different communication rates.
As a result, the node set $\verts$ of $\graph$ is partitioned to yield $M$ cluster sub-graphs, i.e., $\{\graph[\verts_p]\}_{p\in [M]}$, and $M^\ast$ inter-cluster sub-graphs, i.e., $\{\graph[\verts^r]\}_{r\in [M^\ast]}$.\footnote{\fred{Please see Section~\ref{sec: Cluster and Inter-Cluster Sub-Graphs} for the precise details regarding the cluster and inter-cluster sub-graphs considered in this work.}}
For each $p\in\verts$, agent $p$ can measure the outputs of its same-cluster neighbors and itself, that is, $\{y_q\}_{q\in(\nbd_p\cap\cluster_p)\cup\{p\}}$, at isolated times defined by the sequence $\{t_k^p\}_{k=0}^\infty$ which satisfies 
\begin{equation}\label{eqn: Timer bounds tau}
    T_1^p \leq t_{k+1}^p - t_k^p \leq T_2^p 
\end{equation}
for all $k\in\ZZ_{\geq 1}$.
Note, the parameters $T_1^p$ and $T_2^p$ should be selected to satisfy $0<T_1^p\leq T_2^p$.
Recall, the assigned cluster of agent $p$ is denoted by $\cluster_p$ (see Section~\ref{sec: Cluster and Inter-Cluster Sub-Graphs}).
If agent $p$ is also a member of inter-cluster $r\in[M^\ast]$, then agent $p$ can also measure the output of its same-$r$-inter-cluster neighbors and itself at isolated times defined by the sequence $\{\mathtt{t}_k^r\}_{k=0}^\infty$ which satisfies 
\begin{equation} \label{eqn: Timer bounds rho}
    T_3^r \leq \mathtt{t}_{k+1}^r - \mathtt{t}_k^r \leq T_4^r
\end{equation}
for all $k\in\ZZ_{\geq 1}$.
Note, the parameters $T_3^r$ and $T_4^r$ should be selected to satisfy $0<T_3^r\leq T_4^r$.
In light of~\eqref{eqn: Timer bounds tau} and~\eqref{eqn: Timer bounds rho}, the rates of output measurements in the cluster of agent $p$ and inter-cluster $r$ can be controlled through the selection of the parameters $T_1^p$, $T_2^p$, $T_3^r$, and $T_4^r$.}

\fred{The objective of this work is to develop a distributed controller for each agent $p\in\verts$ that yields consensus in the agents' states, that is, 
\begin{equation*}
    x_p = x_q
\end{equation*}
for all $p,q\in\verts$ with stable behavior and robustness.}
In addition, for each $p\in\verts$, the controller of agent $p$ should only utilize intermittent measurements of the outputs in $\{y_q\}_{q\in\nbd_p\cup\{p\}}$, which may be obtained asynchronously.
Such a control strategy better leverages limited resources relative to continuous communication and/or sensing alternatives, readily integrates in digital hardware, and accommodates the natural asynchronous flow of information within MASs.   

\section{\sean{Hybrid System Modeling}} \label{sec: Modeling}
\fred{
Within this section, we develop the closed-loop ensemble hybrid system under our proposed distributed consensus controller.
We begin by introducing the timer mechanisms responsible for the generation of the sequences $\{t_k^p\}_{k=0}^\infty$ and $\{\mathtt{t}_k^r\}_{k=0}^\infty$ discussed in the previous section.
Two different hybrid estimators are introduced next, which are the foundation of our distributed consensus controller.
The final sub-section presents auxiliary error systems and ensemble variables that support the derivation of the closed-loop ensemble hybrid system.}

\subsection{\sean{Timer Mechanisms}}
\fred{Let $0<T_1^p\leq T_2^p$ be user-defined constants for each $p\in\verts$, where $T_1^p$ can be used to account for the sampling rate of the slowest sensor of agent $p$ and $T_2^p$ can be used to account for the MATI of agent $p$.
Furthermore, let $\tau_p\in[0,T_2^p]$ denote the time kept by a timer mechanism belonging to agent $p$ whose evolution is modeled by the hybrid system}
\begin{equation} \label{eqn: tau timer p}
\begin{aligned}
    \dot{\tau}_p &= -1, & \tau_p &\in [0,T_2^p]\\
    \tau_p^+ &\in[T_1^p,T_2^p], & \tau_p &=0
\end{aligned}
\end{equation}
with initial condition $\tau_p(0,0)\in [0,T_2^p]$.\footnote{Examples that leverage the timer mechanism in~\eqref{eqn: tau timer p} can be found in~\cite{Phillips.Sanfelice2019a}, \cite{Goebel.Sanfelice.ea2012}, and~\cite{Zegers.Phillips.ea2024}.}
\fred{The hybrid dynamics in~\eqref{eqn: tau timer p} and the onboard clock of agent $p$ enable the construction of increasing sequences of event times by using the following procedure: 1) set $t_k^p$ for $k=0$ to the initial time given by the onboard clock of agent $p$, 2) increment $k$ by one and allow the timer of agent $p$ to flow using the flow equation in~\eqref{eqn: tau timer p} and the initial condition $\tau_p(0,0)\in [0,T_2^p]$ (during this step, the value of $\tau_p$ is made to decrease at a rate of $-1$ second/second from its initial value), 3) once $\tau_p=0$, set $t_k^p$ for $k=1$ to the current time kept by the onboard clock of agent $p$ and reset $\tau_p$ to any element in $[T_1^p,T_2^p]$ (that is, allow the timer of agent $p$ to jump according to the jump inclusion in~\eqref{eqn: tau timer p}), 4) increment $k$ by one and allow the timer of agent $p$ to flow using the flow equation in~\eqref{eqn: tau timer p} and the value of $\tau_p$ after the jump from Step 3), 5) repeat Steps 3) and 4) \textit{in aeternum}.
The result of this procedure is an increasing sequence of event times, where the event time $t_k^p$ represents the $k^\text{th}$ instant $\tau_p=0$ for agent $p$.
Note, assuming the onboard clock of agent $p$ increases at a constant rate of 1 second/second, the sequence of event times generated by~\eqref{eqn: tau timer p} and the aforementioned procedure satisfies~\eqref{eqn: Timer bounds tau} for all $k\in\ZZ_{\geq 1}$.}

Likewise, for every inter-cluster $r\in[M^\ast]$, let $0<T_3^r\leq T_4^r$ be user-defined constants \fred{playing a similar role to those of $T_1^p$ and $T_2^p$}, and let $\rho_r\in[0,T_4^r]$ \fred{denote the time kept by a timer mechanism assigned to inter-cluster $r$} that evolves according to the hybrid system  
\begin{equation} \label{eqn: rho timer r}
    \begin{aligned}
        \dot{\rho}_r &= -1, & \rho_r &\in [0,T_4^r]\\
        \rho_r^+ &\in[T_3^r,T_4^r], & \rho_r &=0
    \end{aligned}
\end{equation}
with initial condition $\rho_r(0,0)\in [0,T_4^r]$.
The hybrid dynamics in~\eqref{eqn: rho timer r} enables the construction of increasing sequences of \fred{event times}, e.g., $\{\mathtt{t}_k^r\}_{k=0}^\infty$ \fred{following a similar procedure to the one described above}, where the event time $\mathtt{t}_k^r$ represents the $k^\text{th}$ instant $\rho_r=0$.
\fred{Note, the event times in $\{\mathtt{t}_k^r\}_{k=0}^\infty$ can be defined by using a common clock available to all agents of inter-cluster $r$.
For simplicity, we assume the onboard clocks of all agents are synchronized, which facilitates the construction of event time sequences for agents and inter-clusters defined relative to a common time standard.
An exemplar result enabling distributed clock synchronization is~\cite{Guarro.Sanfelice2023}.
Note, assuming the onboard clocks of the agents are synchronized and increase at a constant rate of 1 second/second, the sequence of event times generated by~\eqref{eqn: rho timer r} and a sequence building procedure similar to the one discussed above satisfies~\eqref{eqn: Timer bounds rho} for all $k\in\ZZ_{\geq 1}$.
A convenient feature of the hybrid system in~\eqref{eqn: tau timer p} is that intermittent event times can be generated by ensuring $0<T_1^p<T_2^p$ holds.
Nonetheless, \eqref{eqn: tau timer p} can produce periodic event times provided $0<T_1^p=T_2^p$ holds.
An analogous comment follows for the event times generated by the hybrid system in~\eqref{eqn: rho timer r}.
The timers modeled by~\eqref{eqn: tau timer p} and~\eqref{eqn: rho timer r} can be treated as independent autonomous systems capable of triggering intermittent and asynchronous actions as defined next.}

\subsection{\sean{Consensus Controller Design}}
\sean{To achieve consensus, we design a first-order hybrid controller that utilizes output measurements from connected agents, when they are available, to drive the state of each agent with dynamics in~\eqref{eqn: Agent p dynamics}.}
For every agent $p\in\verts$ and inter-cluster $r\in[M^\ast]$, let $\eta_p\in\RR^m$ and $\zeta_{pr}\in\RR^m$ be auxiliary variables that evolve according to
\begin{equation} \label{eqn: eta_p dynamics}
\begin{aligned}
	\dot{\eta}_p &= K_{\eta,p}\eta_p, & \tau_p &\in[0,T_2^p]\\
	\eta_p^+ &= \sum_{q\in\nbd_p\cap\cluster_p}(y_q - y_p), & \tau_p &= 0
\end{aligned}
\end{equation}
and
\begin{equation} \label{eqn: zeta_pr dynamics}
\begin{aligned}
	\dot{\zeta}_{pr} &= K_{\zeta,pr}\zeta_{pr}, & \rho_r &\in[0,T_4^r]\\
	\zeta_{pr}^+ &= \sum_{q\in\nbd_p\cap\verts^r}(y_q - y_p), & \rho_r &=0,
\end{aligned}
\end{equation}
respectively, where $K_{\eta,p},K_{\zeta,pr}\in\RR^{m\times m}$ are user-defined matrices.
\fred{The systems in~\eqref{eqn: eta_p dynamics} and~\eqref{eqn: zeta_pr dynamics} model estimators belonging to agent $p$ that are intermittently reset with output information during certain jumps and allowed to evolve in an open-loop fashion during flows.
In particular, when a jump is caused by $\tau_p=0$, the value of $\eta_p$ is reset with the value of a consensus-like term computed using only output information from the same-cluster neighbors of agent $p$, as defined by the jump equation in~\eqref{eqn: eta_p dynamics}.
When $\tau_p\in [0,T_2^p]$, the flow equation (i.e., LTI system) in~\eqref{eqn: eta_p dynamics} defines an open-loop trajectory, that is, a trajectory not influenced by output feedback.
Similarly, when a jump is caused by $\rho_r=0$, the value of $\zeta_{pr}$ is reset with the value of a consensus-like term computed using only output information from the neighbors of agent $p$ in inter-cluster $r$, as defined by the jump equation in~\eqref{eqn: zeta_pr dynamics}.
When $\rho_r\in [0,T_4^r]$, the flow equation (i.e., LTI system) in~\eqref{eqn: zeta_pr dynamics} defines an open-loop trajectory.
Since the timers modeled by~\eqref{eqn: tau timer p} and~\eqref{eqn: rho timer r} are independent, the resets experienced by $\eta_p$ and $\zeta_{pr}$ may be asynchronous.
It is worth noting that the hybrid dynamics in~\eqref{eqn: eta_p dynamics} and~\eqref{eqn: zeta_pr dynamics} for every $p\in\verts$ and $r\in[M^\ast]$ induce information flows between agents that are intermittent, asynchronous, and directed, which are taking place over a static, connected, and undirected graph.
Moreover, at consensus, i.e., when $x_p = x_q$ for all $p,q\in\verts$, the variables $\eta_p$ and $\zeta_{pr}$ both equal $0_n$ given~\eqref{eqn: Agent p dynamics}, \eqref{eqn: eta_p dynamics}, and~\eqref{eqn: zeta_pr dynamics}.
It is also worth noting that, even though the estimators values $\eta_p$ and $\zeta_{pr}$ evolve in an open-loop fashion during flows, the timer parameters $T_2^p$ and $T_4^r$ bound the period of respective flow and will be selected to ensure the stability of the closed-loop system.} 

Given a user-defined matrix $K_u\in\RR^{d\times m}$ \fred{and the estimates produced by the estimators in~\eqref{eqn: eta_p dynamics} and~\eqref{eqn: zeta_pr dynamics}}, the controller of agent $p$ is \fred{designed as}
\begin{equation} \label{eqn: agent controller}
    u_p \coloneqq K_u\Bigg(\eta_p + \sum_{r\in[M^\ast]} \zeta_{pr}\Bigg)\in\RR^d.
\end{equation}
By construction, the controller in~\eqref{eqn: agent controller} is distributed as it only requires output information from neighboring agents and the implementing agent itself.
\fred{The} output measurements \fred{can be} obtained intermittently provided $T_1^p<T_2^p$ and $T_3^r<T_4^r$ for all \fred{inter-clusters} $r\in[M^\ast]$.
Furthermore, the output measurements belonging to agents in $\nbd_p\cap\cluster_p$, $\nbd_p\cap\verts^r$, and $\nbd_p\cap\verts^s$, \fred{where} $r\neq s\in[M^\ast]$, may be obtained asynchronously \fred{given the independence of the timer mechanisms.}
Note, if agent $p$ is not a member of inter-cluster $r$ (i.e., $p\notin\verts^r$), $\zeta_{pr}\equiv 0_m$.

\subsection{\sean{Hybrid System Modeling}}
To facilitate the derivation of the closed-loop ensemble hybrid system, \fred{we will leverage the errors systems} 
\begin{align}
    \tilde{\eta}_p &\coloneqq \eta_p - \sum_{q\in\nbd_p\cap\cluster_p}(y_q - y_p)\in\mathbb{R}^m, \label{eqn: etaTilde_p} \\
    \tilde{\zeta}_{pr} &\coloneqq \zeta_{pr} - \sum_{q\in\nbd_p\cap\verts^r}(y_q - y_p)\in\mathbb{R}^m. \label{eqn: zetaTilde_pr}
\end{align}
\fred{Furthermore,} we provide the following notation to aid the writing of concise expressions.
Let
\begin{equation*}
\begin{aligned}
    x &\coloneqq (x_p)_{p\in\verts}\in\RR^{nN}, & \eta &\coloneqq (\eta_p)_{p\in\verts}\in\RR^{mN}, & \zeta_r &\coloneqq (\zeta_{pr})_{p\in\verts}\in\RR^{mN}, & \zeta &\coloneqq (\zeta_r)_{r\in[M^\ast]}\in\RR^{mNM^\ast}, \\
    \tilde{\eta} &\coloneqq (\tilde{\eta}_p)_{p\in\verts}\in\RR^{mN}, & \tilde{\zeta}_r &\coloneqq (\tilde{\zeta}_{pr})_{p\in\verts}\in\RR^{mN}, & \tilde{\zeta} &\coloneqq (\tilde{\zeta}_r)_{r\in[M^\ast]}\in\RR^{mNM^\ast}, & \tau&\coloneqq (\tau_p)_{p\in\verts}\in\RR^N, \\
    \rho &\coloneqq (\rho_r)_{r\in[M^\ast]}\in\RR^{M^\ast}.
\end{aligned}
\end{equation*}
The substitution of the output equation in~\eqref{eqn: Agent p dynamics} and~\eqref{eqn: agent controller}--\eqref{eqn: zetaTilde_pr} into the differential equation in~\eqref{eqn: Agent p dynamics} yields
\begin{equation} \label{eqn: x_pDot}
    \dot{x}_p = Ax_p + BK_u H\sum_{q\in\nbd_p}(x_q - x_p) + BK_u \sum_{r\in[M^\ast]}\tilde{\zeta}_{pr} + BK_u \tilde{\eta}_p,
\end{equation}
where the summation term over the neighbor set of agent $p$ is due to~\eqref{eqn: neighbor set identity}.
Substituting~\eqref{eqn: x_pDot} into the time derivative of $x$ for each $p\in\verts$ yields
\begin{equation} \label{eqn: xDot}
    \dot{x} = ((I_N\otimes A)- (\lap\otimes BK_u H))x + (I_N\otimes BK_u)\tilde{\eta} + (I_N\otimes BK_u)(1_{M^\ast}^\top \otimes I_{mN})\tilde{\zeta}.
\end{equation}
Recall the objective\fred{:} drive $x$ from any initial condition to a configuration where $x_p = x_q$ for all distinct $p,q\in\verts$.
Since any configuration $x$ can be decomposed into orthogonal components, where $x = x^\Vert + x^\perp$, $x^\Vert \coloneqq ((1_N 1_N^\top)/N\otimes I_n)x$ is the part of $x$ in consensus, and $x^\perp \coloneqq (\sproj\otimes I_n)x$ is the part of $x$ in dissensus, one can achieve the objective by driving $\Vert x^\perp\Vert$ to $0$.
Naturally, we attempted to produce a GES result using the dynamics of $x^\perp$.
However, in this particular setting, such an exercise was unfruitful because of the interaction between the non-trivial kernel of \fred{the graph Laplacian matrix} $\lap$ and the linear time-invariant dynamics in~\eqref{eqn: Agent p dynamics}.

The use of $\sproj\lap = \lap\sproj$ and the substitution of~\eqref{eqn: xDot} into the time derivative of $x^\perp$ yields a linear ordinary differential equation (ODE) that depends on the variable $(x^\perp,\tilde{\eta},\tilde{\zeta})$.
The matrix that pre-multiplies $x^\perp$ cannot, in general, be made into a stability matrix through the selection of $K_u$, unless $A$ itself is a stability matrix.
Fortunately, Lemma~\ref{lemma: L and S Identities} enables the succeeding result, which motivates the analysis of an alternative ODE.
\begin{lemma} \label{lemma: Equivalent Consensus Metric}
    For any $z\in\RR^{nN}$, $\Vert z^\perp \Vert = \Vert (\WV^\top\otimes I_n) z\Vert$.
    \hfill$\triangle$
\end{lemma}
\begin{proof}
    By Lemma~\ref{lemma: L and S Identities}, $\sproj = \WV\WV^\top$.
   Moreover, $\sproj = \sproj^\top$ by construction, and $\sproj^2 = \sproj$ since $\sproj$ is a projection.
    Therefore,
    \begin{equation*}
    \begin{aligned}
        \Vert z^\perp \Vert^2 &= \Vert (\sproj\otimes I_n)z \Vert^2 = z^\top (\sproj\otimes I_n)^\top (\sproj\otimes I_n) z = z^\top (\sproj^\top\otimes I_n) (\sproj\otimes I_n) z = z^\top (\sproj^\top\sproj\otimes I_n) z \\
        &= z^\top (\sproj^2\otimes I_n) z = z^\top (\sproj\otimes I_n) z = z^\top (\WV\WV^\top\otimes I_n) z = z^\top (\WV\otimes I_n)(\WV^\top\otimes I_n) z = \Vert (\WV^\top\otimes I_n) z \Vert^2,
    \end{aligned}
    \end{equation*}
    and the desired result follows.
\end{proof}

Let $x^\circ\coloneqq(\WV^\top \otimes I_n)x\in\RR^{n(N-1)}$, and recall that consensus in $\{x_p\}_{p\in\verts}$ is achieved whenever a solution of~\eqref{eqn: xDot} converges to the agreement subspace defined by $\mathtt{A}\coloneqq \{x\in\RR^{nN}\colon\forall_{p,q\in\verts} \ x_p = x_q\}$.
Since the distance of an arbitrary configuration $x$ to the agreement subspace $\mathtt{A}$ is quantified by $\Vert x^\perp \Vert$ and $\Vert x^\perp \Vert = \Vert x^\circ \Vert$ by Lemma~\ref{lemma: Equivalent Consensus Metric}, $\Vert x^\circ \Vert$ is an alternative metric for consensus.
Therefore, $x_p=x_q$ for all distinct $p,q\in\verts$ if and only if $\Vert x^\circ\Vert = 0$.

The substitution of both identities in~\eqref{eqn: Laplacian Identity} and the ODE in~\eqref{eqn: xDot} into the time derivative of $x^\circ$ yields\footnote{We remind the reader that, under Lemma~\ref{lemma: L and S Identities}, $\lap = \WV\WD\WV^\top$ with $\WD$ being a diagonal positive definite matrix.}
\begin{equation} \label{eqn: xCircDot}
    \dot{x}^\circ = ((I_{N-1}\otimes A)- (\WD\otimes BK_u H))x^\circ + (\WV^\top\otimes BK_u)\tilde{\eta} + (\WV^\top\otimes BK_u)(1_{M^\ast}^\top \otimes I_{mN})\tilde{\zeta}.
\end{equation}
Observe, the matrix $(I_{N-1}\otimes A)- (\WD\otimes BK_u H)$ can be made stable through the selection of $K_u$, independent of the stability of $A$, provided $(A,B)$ is controllable, $(A,H)$ is observable, and $K_uH\neq 0_{d\times n}$ given~\cite[Lemma 2.2 \& Theorem 2.6]{Rodrigues2022}.\footnote{Let $\lambda_1>0$ denote the smallest eigenvalue of $\WD$.
Using a partial order on the space of positive semi definite matrices, one can show that $(I_{N-1}\otimes A)- (\WD\otimes BK_u H)$ is a stability matrix provided $A - \lambda_1 BK_u H$ is a stability matrix.
Such a matrix $K_u$ can be computed using\cite[Theorem 2.6]{Rodrigues2022}.  
}
Consider the block diagonal matrix $\mathtt{e}\coloneqq \diag{\mathtt{e}_1,\mathtt{e}_2,...,\mathtt{e}_N}\in\RR^{N^2 \times N}$, where $\mathtt{e}_k$ denotes the $k^\text{th}$ standard basis vector in $\RR^N$.
Substituting $\lap = \lap\sproj$, \eqref{eqn: S Identity}, the output equation in~\eqref{eqn: Agent p dynamics}, \eqref{eqn: etaTilde_p}, and $x^\circ\coloneqq(\WV^\top \otimes I_n)x$ into $\tilde{\eta}$ for every $p\in\verts$ yields
\begin{equation} \label{eqn: etaTilde}
    \tilde{\eta} = \eta + \CC x^\circ,
\end{equation}
where $\lap_0\coloneqq \diag{\lap[\cluster_1],\lap[\cluster_2],...,\lap[\cluster_N]}\in\RR^{N^2 \times N^2}$ and $\CC\coloneqq (I_N\otimes H)(\mathtt{e}^\top\lap_0(I_N\otimes\WV)\otimes I_n)(1_N\otimes I_{n(N-1)})\in\RR^{mN\times n(N-1)}$.
The substitution of the flow equation in~\eqref{eqn: eta_p dynamics} for each $p\in\verts$, \eqref{eqn: xCircDot}, and \eqref{eqn: etaTilde} into the time derivative of~\eqref{eqn: etaTilde} yields
\begin{equation} \label{eqn: etaTildeDot}
    \dot{\tilde{\eta}} = (\CC((I_{N-1}\otimes A)- (\WD\otimes BK_u H)) - K_{\eta}\CC)x^\circ + (\CC(\WV^\top\otimes BK_u) + K_{\eta})\tilde{\eta} + \CC(\WV^\top\otimes BK_u)(1_{M^\ast}^\top \otimes I_{mN})\tilde{\zeta}
\end{equation}
with $K_{\eta}\coloneqq\diag{K_{\eta,1},K_{\eta,2},...,K_{\eta,N}}\in\RR^{mN\times mN}$.
Further, the substitution of $\lap = \lap\sproj$, \eqref{eqn: S Identity}, the output equation in~\eqref{eqn: Agent p dynamics}, \eqref{eqn: zetaTilde_pr}, and $x^\circ\coloneqq(\WV^\top \otimes I_n)x$ into $\tilde{\zeta}$ for every $p\in\verts$ and $r\in[M^\ast]$ yields
\begin{equation} \label{eqn: zeta_rTilde}
    \tilde{\zeta}_r = \zeta_r + \mathsf{I}_r x^\circ,
\end{equation}
where $\II_r \coloneqq (I_N\otimes H)(\mathtt{e}^\top(I_N \otimes \lap[\verts^r]\WV)\otimes I_n)(1_N\otimes I_{n(N-1)})\in\RR^{mN\times n(N-1)}$.
Substituting~\eqref{eqn: zeta_rTilde} for every $r\in[M^\ast]$ into $\tilde{\zeta}$ yields
\begin{equation} \label{eqn: zetaTilde}
    \tilde{\zeta} = \zeta + \II x^\circ,
\end{equation}
where $\II\coloneqq \diag{\mathsf{I}_1,\mathsf{I}_2,...,\mathsf{I}_{M^\ast}}(1_{M^\ast}\otimes I_{n(N-1)})$.
The substitution of the flow equation in~\eqref{eqn: zeta_pr dynamics} for every $p\in\verts$ and $r\in[M^\ast]$, \eqref{eqn: xCircDot}, and~\eqref{eqn: zetaTilde} into the time derivative of~\eqref{eqn: zetaTilde} yields
\begin{equation} \label{eqn: zetaTildeDot}
    \dot{\tilde{\zeta}} = (\II((I_{N-1}\otimes A)- (\WD\otimes BK_u H)) - K_{\zeta}\II)x^\circ + \II(\WV^\top\otimes BK_u)\tilde{\eta} + (\II(\WV^\top\otimes BK_u)(1_{M^\ast}^\top \otimes I_{mN}) + K_{\zeta})\tilde{\zeta},
\end{equation}
where $K_{\zeta,r}\coloneqq\diag{K_{\zeta,1r},K_{\zeta,2r},...,K_{\zeta,Nr}}\in\mathbb{R}^{mN\times mN}$ and $K_{\zeta}\coloneqq\diag{K_{\zeta,1},K_{\zeta,2},...,K_{\zeta,M^{\ast}}}\in\mathbb{R}^{mNM^{\ast}\times mNM^{\ast}}$.

Let $z\coloneqq (x^\circ,\tilde{\eta},\tilde{\zeta})\in\RR^{n(N-1)+mN+mNM^\ast}$.
By leveraging the ODEs in~\eqref{eqn: xCircDot}, \eqref{eqn: etaTildeDot}, and~\eqref{eqn: zetaTildeDot}, a dynamical system can be developed for the variable $z$ during flows, namely,
\begin{equation} \label{eqn: z dynamics}
    \dot{z} = \fmat z,
\end{equation}
where
\begin{equation*}
    \fmat \coloneqq
    \left[
    \begin{array}{ccc}
        \fmat_{11} & \fmat_{12}  & \fmat_{13} \\
        \CC\fmat_{11} - K_{\eta}\CC &  \CC\fmat_{12} + K_{\eta}& \CC\fmat_{13} \\
        \II\fmat_{11} - K_{\zeta}\II &  \mathsf{I}\fmat_{12} & \mathsf{I}\fmat_{13} + K_{\zeta}
    \end{array}
    \right], \quad \quad
    \begin{aligned}
    \fmat_{11} &\coloneqq (I_{N-1}\otimes A)- (\WD\otimes BK_u H), \\ 
    \fmat_{12} & \coloneqq \WV^\top\otimes BK_u, \\
    \fmat_{13} &\coloneqq (\WV^\top\otimes BK_u)(1_{M^\ast}^\top \otimes I_{mN}).
    \end{aligned}
\end{equation*}
Let $\hyb$ denote the closed-loop hybrid system for the ensemble.
The state variable and state space of $\hyb$ are denoted by $\xi\coloneqq (z,\tau,\rho)\in\statespc$ and $\statespc\coloneqq \RR^{n(N-1)}\times\RR^{mN}\times\RR^{mNM^\ast}\times\tspc\times\rspc$, respectively.
\fred{Observe,} $\tspc\coloneqq [0,T_2^1]\times[0,T_2^2]\times...\times[0,T_2^N]$ and $\rspc\coloneqq [0,T_4^1]\times[0,T_4^2]\times...\times[0,T_2^{M^\ast}]$.
For every $p\in\verts$ and $r\in[M^\ast]$, let $C_{\tau,p} \coloneqq \{\xi\in\statespc\colon \tau_p\in[0,T_2^p]\}$ and $C_{\rho,r} \coloneqq \{\xi\in\statespc\colon \rho_r\in[0,T_4^r]\}$.
The flow set of $\hyb$ is
\begin{equation} \label{eqn: flow set}
    C \coloneqq \big(\cap_{p\in\verts} C_{\tau,p} \big) \cap \big(\cap_{r\in[M^\ast]} C_{\rho,r} \big).
\end{equation}
The differential equation governing the flows of $\hyb$ is $\dot{\xi} = f(\xi)$, where the single-valued flow map $f\colon\statespc\to\statespc$ is
\begin{equation} \label{eqn: flow map}
    f(\xi) \coloneqq (\fmat z, -1_N, -1_{M^\ast}).
\end{equation}
The flow map in~\eqref{eqn: flow map} follows from the substitution of the flow equation in~\eqref{eqn: tau timer p} for each $p\in\verts$, the flow equation in~\eqref{eqn: rho timer r} for each $r\in[M^\ast]$, and the dynamics of $z$ in~\eqref{eqn: z dynamics} into the time derivative of $\xi$.
For each $p\in\verts$ and $r\in[M^\ast]$, let $D_{\tau,p} \coloneqq \{\xi\in\statespc\colon \tau_p=0\}$ and $D_{\rho,r} \coloneqq \{\xi\in\statespc\colon \rho_r=0\}$.
The jump set of $\hyb$ is
\begin{equation} \label{eqn: jump set}
    D \coloneqq \big(\cup_{p\in\verts} D_{\tau,p} \big) \cup \big(\cup_{r\in[M^\ast]} D_{\rho,r} \big).
\end{equation}
The \fred{difference} inclusion governing the jumps of $\hyb$ is $\xi^+ \in G(\xi)$, where the set-valued jump map $G\colon\statespc\rightrightarrows\statespc$ is
\begin{equation} \label{eqn: jump map}
\begin{aligned}
    G(\xi) &\coloneqq \{G_{\tau,p}(\xi)\colon\xi\in D_{\tau,p}\text{ for some }p\in\verts\} \cup \{G_{\rho,r}(\xi)\colon\xi\in D_{\rho,r}\text{ for some }r\in[M^\ast]\}, \\
    G_{\tau,p}(\xi) &\coloneqq 
    \left[ \begin{array}{c}
    x^\circ \\
    \left[\tilde{\eta}_1^\top,...,\tilde{\eta}_{p-1}^\top,0_m^\top,\tilde{\eta}_{p+1}^\top,...,\tilde{\eta}_N^\top\right]^\top \\
    \tilde{\zeta} \\
    \left[\tau_1,...,\tau_{p-1},[T_1^p,T_2^p],\tau_{p+1},...,\tau_N\right]^\top \\
    \rho
    \end{array} \right], \quad
    G_{\rho,r}(\xi) \coloneqq 
    \left[ \begin{array}{c}
    x^\circ \\
    \tilde{\eta} \\
    \left[\tilde{\zeta}_1^\top,...,\tilde{\zeta}_{r-1}^\top,0_{mN}^\top,\tilde{\zeta}_{r+1}^\top,...,\tilde{\zeta}_{M^\ast}^\top\right]^\top \\
    \tau \\
    \left[\rho_1,...,\rho_{r-1},[T_3^r,T_4^r],\rho_{r+1},...,\rho_{M^\ast}\right]^\top
    \end{array} \right].
\end{aligned}
\end{equation}
The jump map in~\eqref{eqn: jump map} is derived by employing the following observations.
Recall from~\eqref{eqn: Agent p dynamics} that, for each $p\in\verts$, $x_p$ evolves continuously, which implies that $x_p^+=x_p$ at jumps.
Therefore, $(x^\circ)^+=x^\circ$ given $x^\circ = (\WV^\top\otimes I_n)x$.
The jump inclusion in~\eqref{eqn: tau timer p} implies $\tau_p^+\in [T_1^p,T_2^p]$ in response to a jump caused by $\tau_p = 0$, and $\tau_p^+ = \tau_p$ in response to a jump not caused by $\tau_p = 0$.
Likewise, the jump inclusion in~\eqref{eqn: rho timer r} implies $\rho_r^+\in [T_3^r,T_4^r]$ in response to a jump caused by $\rho_r = 0$, and $\rho_r^+ = \rho_r$ in response to a jump not caused by $\rho_r = 0$.
The jump equation in~\eqref{eqn: eta_p dynamics} and the definition in~\eqref{eqn: etaTilde_p} imply $\tilde{\eta}_p^+=0_m$ whenever a jump occurs in response to $\tau_p = 0$, and $\tilde{\eta}_p^+=\tilde{\eta}_p$ otherwise.
Similarly, the jump equation in~\eqref{eqn: zeta_pr dynamics} and the definition in~\eqref{eqn: zetaTilde_pr} imply $\tilde{\zeta}_{pr}^+=0_m$ for all $p\in\verts^r$ whenever a jump occurs in response to $\rho_r = 0$, and $\tilde{\zeta}_{pr}^+=\tilde{\zeta}_{pr}$ for all $p\in\verts^r$ otherwise.

The solutions of the hybrid system $\hyb$ with data $(C,f,D,G)$ describe the behavior of the ensemble.
Consequently, the MAS accomplishes consensus in the agents' states provided the set
\begin{equation} \label{eqn: attractor}
    \attr \coloneqq \big\{\xi\in\statespc\colon (x^\circ,\tilde{\eta},\tilde{\zeta}) = (0_{n(N-1)},0_{mN},0_{mNM^\ast})\big\}
\end{equation}
is GES.
Therefore, the consensus problem can be transformed into a set stabilization problem for dynamical systems, which is the main focus of this work. 
In addition, since the set $\attr$ can be alternatively written as $\attr=\{\xi\in\statespc\colon \Vert z \Vert = 0\}$, $\xi^\prime\in\attr$ if and only if $\xi^\prime = (\mathbf{0},\tau^\prime,\rho^\prime)$, $\tau^\prime\in\tspc$, and $\rho^\prime\in\rspc$, where $\mathbf{0}$ denotes a zero vector of appropriate dimension.
As a result, for any $\xi\in\statespc$,
\begin{equation} \label{eqn: distance to desired set}
    \vert \xi \vert_{\attr} = \inf\{\Vert \xi - \xi^\prime\Vert\colon\xi^\prime\in\attr\} = \Vert z \Vert.
\end{equation}

Under the construction of $\hyb$, the flow set $C$ and jump set $D$ are closed.
Additionally, the flow map $f$ is continuous.
The jump map $G$ is outer semi-continuous and locally bounded.
As a result, $\hyb$ satisfies the hybrid basic conditions~\cite[Assumption 6.5]{Goebel.Sanfelice.ea2012}, and~\cite[Theorem 6.8]{Goebel.Sanfelice.ea2012} implies $\hyb$ is nominally well-posed.

\section{Stability Analysis} \label{sec: Stability Analysis}
Before demonstrating $\attr$ is GES for the hybrid system $\hyb$, maximal solutions must first be shown to be complete so that the quantity $\vert\phi(t,j)\vert_{\attr}$ is well-defined for arbitrarily large $t+j$.
Two useful inequalities relating the continuous-time variable $t$ and discrete-time variable $j$ are also provided, which rely on the following constants:
\begin{equation*}
    T_{\min} \coloneqq \min \{T_1^p\colon p\in\verts\}\cup\{T_3^r\colon r\in[M^\ast]\}, \quad T_{\max} \coloneqq \max \{T_2^p\colon p\in\verts\}\cup\{T_4^r\colon r\in[M^\ast]\}.
\end{equation*}
\begin{lemma} \label{lemma: Completeness and (t,j) Bounds}
    For every maximal solution $\phi$ of the hybrid system $\hyb$ with data $(C,f,D,G)$, the following items hold:
    \begin{enumerate}
        \item $\phi$ is complete \fred{and non-Zeno;}
        \item The continuous-time $t$ can be bounded by functions of the discrete-time $j$, i.e.,
        \begin{equation} \label{eqn: (t,j) inequality}
           \left(\frac{j}{N+M^\ast} - 1\right)T_{\min} \leq t \leq \frac{j}{N+M^\ast} T_{\max}
        \end{equation}
        for all $(t,j)\in\dom\phi$.
        \hfill$\triangle$
    \end{enumerate}
\end{lemma}
\begin{proof}
    See Appendix~\ref{appen: completeness and (t,j) bounds}.
\end{proof}

To streamline the presentation of the main result, consider the following items.
Let 
\begin{equation*}
    T_2 \coloneqq (T_2^p)_{p\in\verts}\in\RR^N, \quad T_4 \coloneqq (T_4^r)_{r\in[M^\ast]}\in\RR^{M^\ast}, \quad T_2^{\max} \coloneqq \max\{T_2^p\colon p\in\verts \}\in\RR_{> 0}.
\end{equation*}
Let $\sigma\in\RR_{>0}$ be a user-defined constant.
Furthermore, let $P_1\in\RR^{n(N-1)\times n(N-1)}$, $P_{2,k}\in\RR^{m\times m}$ for each $k\in\verts$, and $P_{3,\ell}\in\RR^{mN\times mN}$ for each $\ell\in[M^\ast]$ be user-defined, symmetric, and positive definite matrices.
We can then use these matrices to define the following block diagonal matrices:
\begin{equation} \label{eqn: P(tau,rho)}
\begin{aligned}
    P_2(\tau) & \coloneqq \diag{P_{2,1}\exp(\sigma\tau_1),P_{2,2}\exp(\sigma\tau_2),...,P_{2,N}\exp(\sigma\tau_N)}\in\RR^{mN\times mN}, \\
    P_3(\rho) & \coloneqq \diag{P_{3,1}\exp(\sigma\rho_1),P_{3,2}\exp(\sigma\rho_2),...,P_{3,M^\ast}\exp(\sigma\rho_{M^\ast})}\in\RR^{mNM^\ast\times mNM^\ast}, \\
    P(\tau,\rho) &\coloneqq \diag{P_1,P_2(\tau),P_3(\rho)}.
\end{aligned}
\end{equation}
Moreover, let 
\begin{equation} \label{eqn: Theorem 1 items}
\begin{aligned}
    \mmat(\tau,\rho) &\coloneqq \fmat^\top P(\tau,\rho) + P(\tau,\rho)\fmat + \dot{P}(\tau,\rho), \quad \mu \coloneqq -\sup\{\lambda_{\max}(\mmat(\tau,\rho))\colon (\tau,\rho)\in\tspc\times\rspc\}, \\
    \alpha_1 &\coloneqq \lambda_{\min}(P(0_N,0_{M^\ast})), \quad \alpha_2\coloneqq \lambda_{\max}(P(T_2,T_4)), \quad \dot{P}(\tau,\rho) = -\sigma\diag{0_{n(N-1)\times n(N-1)}, P_2(\tau), P_3(\rho)}.
\end{aligned}
\end{equation}
By definition, the matrix $\mmat(\tau,\rho)$ is symmetric.
If $\mmat(\tau,\rho)$ satisfies $\mmat(\tau,\rho)<\mathbf{0}$ for all $(\tau,\rho)\in\tspc\times\rspc$, then the constant $\mu$ is positive.
Further, $\alpha_1\in\RR_{>0}$ and $\alpha_2\in\RR_{>0}$ since $P(\tau,\rho)$ is positive definite.

\begin{theorem} \label{thm: A is GES} 
If there exist timer parameters $0<T_1^p\leq T_2^p$ for every $p\in\verts$ and $0< T_3^r\leq T_4^r$ for every $r\in[M^\ast]$, a constant $\sigma$, controller matrix $K_u$, estimator matrices $K_{\eta,p}$ for each $p\in\verts$, estimator matrices $K_{\zeta,pr}$ for every $(p,r)\in\verts\times[M^\ast]$, and symmetric, positive definite matrices $P_1$, $P_2(\tau)$, and $P_3(\rho)$ that satisfy the inequality $\mmat(\tau,\rho) < \mathbf{0}$ for all $(\tau,\rho)\in\tspc\times\rspc$, then the set $\attr$ in~\eqref{eqn: attractor} is GES for the hybrid system $\hyb$ with data $(C,f,D,G)$.
In particular, for every maximal solution $\phi$ of $\hyb$,
\begin{equation} \label{eqn: GES bound}
    \vert \phi(t,j) \vert_{\attr} \leq \kappa\exp(-\alpha(t+j))\vert \phi(0,0)\vert
\end{equation}
for all $(t,j)\in\dom\phi$, where, for some constant $\varepsilon\in(0,1)$, 
\begin{equation*} 
    \kappa \coloneqq \sqrt{\frac{\alpha_2}{\alpha_1}\exp\left(\frac{(1-\varepsilon)\mu T_{\min}}{\alpha_2}\right)}\in\RR_{>0}, \quad \alpha \coloneqq \frac{1}{2}\min\left\{\frac{\varepsilon\mu}{\alpha_2},\frac{(1-\varepsilon)\mu T_{\min}}{\alpha_2(N+M^{\ast})}\right\}\in\RR_{>0},
\end{equation*}
and $\mu\in\RR_{>0}$.
Furthermore, the controllers $\{u_p(\phi(t,j))\}_{p\in\verts}$ in~\eqref{eqn: agent controller} are bounded for all $(t,j)\in\dom\phi$.
\hfill$\triangle$
\end{theorem}
\begin{proof}
Consider the Lyapunov function candidate
\begin{equation} \label{eqn: V}
    V\colon\statespc\to\RR_{\geq 0}\colon \xi \mapsto
    z^\top P(\tau,\rho) z.
\end{equation}
Using $\vert\xi\vert_{\attr}=\Vert z\Vert$ given~\eqref{eqn: distance to desired set} and the definitions of $\alpha_1$ and $\alpha_2$ in~\eqref{eqn: Theorem 1 items}, the Lyapunov function candidate $V(\xi)$ can be bounded as
\begin{equation} \label{eqn: V Bounds}
    \alpha_1\vert\xi\vert_{\attr}^2 \leq V(\xi)\leq \alpha_2\vert\xi\vert_{\attr}^2.
\end{equation}
When $\xi\in C$, the change in $V(\xi)$ is computed using $\dot{V}(\xi)=\langle\nabla V(\xi),f(\xi)\rangle$ given that $V(\xi)$ is continuously differentiable in $\xi$.
Substituting the flow map in~\eqref{eqn: flow map} into the time derivative of~\eqref{eqn: V} yields
\begin{equation} \label{eqn: VDot 1} 
    \dot{V}(\xi) = 2z^\top P(\tau,\rho) \fmat z + z^\top \dot{P}(\tau,\rho) z = z^\top \big(\fmat^\top P(\tau,\rho) + P(\tau,\rho)\fmat + \dot{P}(\tau,\rho)\big) z = z^\top \mmat(\tau,\rho) z.
\end{equation}
Since $\mmat(\tau,\rho)$ is negative definite for all $(\tau,\rho)\in\tspc\times\rspc$ by the hypothesis, the constant $\mu$ is positive.
Using the definition of $\mu$ in~\eqref{eqn: Theorem 1 items}, $\vert\xi\vert_{\attr}=\Vert z\Vert$, and $V(\xi)\leq \alpha_2\vert\xi\vert_{\attr}^2$ from~\eqref{eqn: V Bounds}, one can bound~\eqref{eqn: VDot 1} as
\begin{equation} \label{eqn: VDot 2}
    \dot{V}(\xi) \leq -\mu\Vert z\Vert^2 = -\mu \vert \xi \vert_{\attr}^2 \leq -\frac{\mu}{\alpha_2} V(\xi).
\end{equation}

When $\xi\in D$ and $g\in G(\xi)$, the change in $V(\xi)$
is computed using $\Delta V(\xi) = V(g)-V(\xi)$.
Suppose $\tau_q=0$ and $\rho_s=0$ for some $q\in\verts$ and $s\in[M^{\ast}]$ without loss of generality.
Hence, the jump map in~\eqref{eqn: jump map} implies that\footnote{The result in~\eqref{eqn: V Jump} also follows if $\tau_q=0$, $\rho_s=0$, or any combination of timers take on a value of $0$.} 
\begin{equation} \label{eqn: V Jump}
    \Delta V(\xi) = (z^+)^\top P(\tau^+,\rho^+)(z^+) - z^\top P(\tau,\rho) z = -\tilde{\eta}_q^\top P_{2,q}\tilde{\eta}_q -\tilde{\zeta}_s^\top P_{3,s}\tilde{\zeta}_s \leq 0
\end{equation}
since $(x^\circ)^+ = x^\circ$ after any jump, $\tilde{\eta}_q^+ = 0_m$ and $\tilde{\eta}_p^+=\tilde{\eta}_p$ for all $p\in\verts\setminus\{q\}$ in response to a jump caused by $\tau_q = 0$, and $\tilde{\zeta}_s^+ = 0_{mN}$ and $\tilde{\zeta}_r^+=\tilde{\zeta}_r$ for all $r\in[M^\ast]\setminus\{s\}$ in response to a jump caused by $\rho_s = 0$.

Next, fix a maximal solution $\phi$ of $\hyb$, select $(t,j)\in\dom\phi$, and let $0= t_0 \leq t_1 \leq ... \leq t_j\leq t$ satisfy
\begin{equation*}
    \dom\phi\bigcap([0,t_j]\times\{ 0,1,...,j-1\})  =\bigcup_{k=1}^j ([t_{k-1},t_k]\times\{k-1\}).
\end{equation*}
For every $k\in\{1,2,...,j\}$ and for almost all $h\in[t_{k-1},t_k]$, $\phi(h,k-1)\in C$.
Furthermore, for every $k\in\{1,2,...,j\}$ and for almost all $h\in[t_{k-1},t_k]$, \eqref{eqn: VDot 2} implies
\begin{equation} \label{eqn: V differential inequality}
    \frac{\dd}{\dd h} V(\phi(h,k-1)) \leq -\frac{\mu}{\alpha_2} V(\phi(h,k-1)).
\end{equation}
Integrating both sides of~\eqref{eqn: V differential inequality}
yields 
\begin{equation} \label{eqn: V solution over flow interval}
    V(\phi(t_k,k-1)) \leq \exp\left(-\frac{\mu}{\alpha_2}(t_k-t_{k-1})\right) V(\phi(t_{k-1},k-1))
\end{equation}
for every $k\in\{1,2,...,j\}$.
Similarly, for each $k\in\{1,2,...,j\}$
with $\phi(t_k,k-1)\in D$, \eqref{eqn: V Jump} implies
\begin{equation} \label{eqn: V jump along solutions}
    V(\phi(t_k,k)) \leq V(\phi(t_k,k-1)).
\end{equation}
By inductively stitching the inequalities in~\eqref{eqn: V solution over flow interval} and~\eqref{eqn: V jump along solutions} along the maximal solution $\phi$, it follows that
\begin{equation} \label{eqn: V GES Bound}
    V(\phi(t,j)) \leq \exp\left(-\frac{\mu}{\alpha_2}t\right) V(\phi(0,0)).
\end{equation}
Using the left inequality in~\eqref{eqn: (t,j) inequality} and the identity $t = \varepsilon t + (1-\varepsilon)t$, one can derive
\begin{equation} \label{eqn: t+j inequality}
    -\frac{\mu}{\alpha_2}t \leq -\min\left\{ \frac{\varepsilon\mu}{\alpha_2},\frac{(1-\varepsilon)\mu T_{\min}}{\alpha_2(N+M^{\ast})}\right\} (t+j) + \frac{(1-\varepsilon)\mu T_{\min}}{\alpha_2},
\end{equation}
which, when substituted into~\eqref{eqn: V GES Bound}, yields
\begin{equation} \label{eqn: V GES Bound 2}
    V(\phi(t,j)) \leq \exp\left(-\min\left\{\frac{\varepsilon\mu}{\alpha_2},\frac{(1-\varepsilon)\mu T_{\min}}{\alpha_2(N+M^{\ast})}\right\}(t+j)\right) \exp\left(\frac{\left(1-\varepsilon\right)\mu T_{\min}}{\alpha_{2}}\right)V\left(\phi\left(0,0\right)\right).
\end{equation}
The application of the inequalities in~\eqref{eqn: V Bounds} on~\eqref{eqn: V GES Bound 2} yields
\begin{equation} \label{eqn: V GES Bound 3}
    \vert\phi(t,j)\vert_{\attr}^2 \leq \frac{\alpha_2}{\alpha_1}\exp\left(-\min\left\{\frac{\varepsilon\mu}{\alpha_2},\frac{(1-\varepsilon)\mu T_{\min}}{\alpha_2(N+M^{\ast})}\right\}(t+j)\right) \exp\left(\frac{\left(1-\varepsilon\right)\mu T_{\min}}{\alpha_{2}}\right)\vert\phi(0,0)\vert_{\attr}^2,
\end{equation}
which leads to the desired bound in~\eqref{eqn: GES bound}.
Therefore, the set $\attr$ is GES since every maximal solution $\phi$ of $\hyb$ is complete by Lemma~\ref{lemma: Completeness and (t,j) Bounds} and $\vert\phi(t,j)\vert_{\attr}$ is bounded according to~\eqref{eqn: GES bound}, with $\kappa,\alpha>0$, for all $(t,j)\in\dom\phi$.

Using~\eqref{eqn: V} and~\eqref{eqn: V GES Bound 2}, one can conclude that $z=z(\phi(t,j))$ is bounded for all $(t,j)\in\dom\phi$.
Thus, $z=(x^\circ,\tilde{\eta},\tilde{\zeta})$ \fred{being bounded implies the disagreement in the states $\{x_p\}_{p\in\verts}$}, $\tilde{\eta}_p$, and $\tilde{\zeta}_{pr}$ are bounded for all $p\in\verts$ and all $r\in[M^{\ast}]$.
\fred{Using the output equation in~\eqref{eqn: Agent p dynamics}, \eqref{eqn: eta_p dynamics}--\eqref{eqn: agent controller}, the bounded flow intervals of $\hyb$, i.e., $t_{j+1} - t_j \leq T_{\max}$ for all $(t_j)_{j=0}^\infty$, and the bounded disagreement in the states $\{x_p\}_{p\in\verts}$, one can demonstrate} $u_p$ is bounded for all $p\in\verts$.
\end{proof}

With respect to Theorem~\ref{thm: A is GES}, the \fred{GES} of $\attr$ for $\hyb$ requires $\mmat(\tau,\rho)<\mathbf{0}$ to hold for all $(\tau,\rho)\in\tspc\times\rspc$.
In other words, $\mmat(\tau,\rho)<\mathbf{0}$ must hold for an infinite number of points, which cannot be exhaustively verified in practice.
\fred{
Hence, we require a practical means by which to verify $\mmat(\tau,\rho)<\mathbf{0}$ holds for all $(\tau,\rho)\in\tspc\times\rspc$, which motivates the following result.}
\begin{corollary} \label{cor: Matrix Inequality}
    If there exist timer parameters $0<T_1^p\leq T_2^p$ for each $p\in\verts$ and $0< T_3^r\leq T_4^r$ for each $r\in[M^\ast]$, a constant $\sigma$, controller matrix $K_u$, estimator matrices $K_{\eta,p}$ for each $p\in\verts$, estimator matrices $K_{\zeta,pr}$ for every $(p,r)\in\verts\times[M^\ast]$, and symmetric, positive definite matrices $P_1$, $P_{2,p}$ for every $p\in\verts$, and $P_{3,r}$ for every $r\in[M^\ast]$ that satisfy $\mmat(0_N,0_{M^\ast}) < \mathbf{0}$ and $\mmat(T_2,T_4) < \mathbf{0}$, then $\mmat(\tau,\rho) < \mathbf{0}$ for all $(\tau,\rho)\in\tspc\times\rspc$.
    \hfill$\triangle$
\end{corollary}
\begin{proof}
\fred{
The following argument is inspired by the proof of Proposition 3.9 in~\cite{Li.Phillips.ea2018}.
For each $p\in\verts$ and $r\in[M^\ast]$, let $\gamma_{\tau,p}\colon [0,T_2^p]\to[0,1]$ and $\gamma_{\rho,r}\colon [0,T_4^r]\to[0,1]$, such that 
\begin{equation} \label{eqn: Convex functions}
    \gamma_{\tau,p}(w) \coloneqq \frac{\exp(\sigma w) - \exp(\sigma T_2^p)}{1-\exp(\sigma T_2^p)}, \quad \gamma_{\rho,r}(w) \coloneqq \frac{\exp(\sigma w) - \exp(\sigma T_4^r)}{1-\exp(\sigma T_4^r)}.
\end{equation}
Let $\epsilon\in[0,1]$ and $\theta\coloneqq(\epsilon,\tau,\rho)\in\Theta$, where $\Theta\coloneqq[0,1]\times\tspc\times\rspc$.
For each $p\in\verts$ and $r\in[M^\ast]$, the functions in~\eqref{eqn: Convex functions} can be manipulated to obtain
\begin{equation} \label{eqn: Convex functions Two}
    \exp(\sigma\tau_p) = \gamma_{\tau,p}(\tau_p) + (1-\gamma_{\tau,p}(\tau_p))\exp(\sigma T_2^p), \quad \exp(\sigma\rho_r) = \gamma_{\rho,r}(\rho_r) + (1-\gamma_{\rho,r}(\rho_r))\exp(\sigma T_4^r).
\end{equation}
One can then construct the following block diagonal matrices:
\begin{equation} \label{eqn: R matrices}
\begin{aligned}
    R_2(\tau) &\coloneqq \diag{\gamma_{\tau,1}(\tau_1),\gamma_{\tau,2}(\tau_2),...,\gamma_{\tau,N}(\tau_N)}\otimes I_m, \quad R_3(\rho) \coloneqq \diag{\gamma_{\rho,1}(\rho_1),\gamma_{\rho,2}(\rho_2),...,\gamma_{\rho,{M^\ast}}(\rho_{M^\ast})}\otimes I_{mN}, \\
    R(\theta) &\coloneqq \diag{(1-\epsilon) I_{n(N-1)},R_2(\tau),R_3(\rho)},
\end{aligned}
\end{equation}
where $R_2(\tau)\in\RR^{mN\times mN}$ and $R_3(\rho)\in\RR^{mNM^\ast\times mNM^\ast}$.
Using the definition of $P(\tau,\rho)$ in~\eqref{eqn: P(tau,rho)}, the equalities in~\eqref{eqn: Convex functions Two}, and the definition of $R(\theta)$ in~\eqref{eqn: R matrices}, one can derive 
\begin{equation} \label{eqn: P and PDot convex forms}
    P(\tau,\rho) = R(\theta)P(0_N,0_{M^{\ast}}) + (I_{\varrho}-R(\theta)) P(T_2,T_4), \quad \dot{P}(\tau,\rho) = R(\theta)\dot{P}(0_N,0_{M^{\ast}}) + (I_{\varrho}-R(\theta))\dot{P}(T_2,T_4),
\end{equation}
where $\varrho \coloneqq n(N-1)+mN+mNM^{\ast}$.
Let $v$ be an arbitrary real-valued vector such that the quadratic form $v^\top\mmat(\tau,\rho) v$ is well-defined.
Using the definition of $\mmat(\tau,\rho)$ in~\eqref{eqn: Theorem 1 items}, it follows that
\begin{equation} \label{eqn: quadratic form identity}
    v^\top\mmat(\tau,\rho) v = v^\top \left(2P(\tau,\rho) \fmat + \dot{P}(\tau,\rho)\right)v.
\end{equation}
The substitution of~\eqref{eqn: P and PDot convex forms} into~\eqref{eqn: quadratic form identity} yields
\begin{equation} \label{eqn: convex quadratic form}
    v^\top\mmat(\tau,\rho) v = v^\top R(\theta)(2P(0_N,0_{M^{\ast}})\fmat + \dot{P}(0_N,0_{M^{\ast}}))v  + v^\top(I_{\varrho} -R(\theta))(2P(T_2,T_4)\fmat + \dot{P}(T_2,T_4))v.
\end{equation}
Since the matrices $P(\tau,\rho)$, $\dot{P}(\tau,\rho)$, $R(\theta)$, and $I_{\varrho} -R(\theta)$ are symmetric, \eqref{eqn: convex quadratic form} can be simplified to
\begin{equation} \label{eqn: M equality}
    v^\top\mmat(\tau,\rho) v = v^\top B_1(\theta)\mmat(0_N,0_{M^{\ast}}) B_1(\theta) v + v^\top B_2(\theta)\mmat(T_2,T_4) B_2(\theta)v,
\end{equation}
where 
\begin{equation*}
\begin{aligned}
    B_{1,2}(\tau) &\coloneqq \text{diag}\left(\sqrt{\gamma_{\tau,1}(\tau_1)},\sqrt{\gamma_{\tau,2}(\tau_2)},...,\sqrt{\gamma_{\tau,N}(\tau_N)}\right)\otimes I_m, \\
    B_{1,3}(\rho) &\coloneqq \text{diag}\left(\sqrt{\gamma_{\rho,1}(\rho_1)},\sqrt{\gamma_{\rho,2}(\rho_2)},...,\sqrt{\gamma_{\rho,{M^\ast}}(\rho_{M^\ast})}\right)\otimes I_{mN}, \\
    B_1(\theta) &\coloneqq \text{diag}\left(\sqrt{1-\epsilon} I_{n(N-1)},B_{1,2}(\tau),B_{1,3}(\rho)\right) \\
    B_{2,2}(\tau) &\coloneqq \text{diag}\left(\sqrt{1-\gamma_{\tau,1}(\tau_1)},\sqrt{1-\gamma_{\tau,2}(\tau_2)},...,\sqrt{1-\gamma_{\tau,N}(\tau_N)}\right)\otimes I_m, \\
    B_{2,3}(\rho) &\coloneqq \text{diag}\left(\sqrt{1-\gamma_{\rho,1}(\rho_1)},\sqrt{1-\gamma_{\rho,2}(\rho_2)},...,\sqrt{1-\gamma_{\rho,{M^\ast}}(\rho_{M^\ast})}\right)\otimes I_{mN}, \\
    B_2(\theta) &\coloneqq \text{diag}\left(\sqrt{\epsilon} I_{n(N-1)},B_{2,2}(\tau),B_{2,3}(\rho)\right).
\end{aligned}
\end{equation*}
Note that $B_1(\theta) v = 0_{\varrho}$ and $B_2(\theta) v = 0_{\varrho}$ cannot hold simultaneously as $B_1(\theta) + B_2(\theta)$ possesses the trivial null space.
Consequently, $\mmat(0_N,0_{M^{\ast}})<\mathbf{0}$, $\mmat(T_2,T_4)<\mathbf{0}$, and~\eqref{eqn: M equality} imply $v^\top\mmat(\tau,\rho) v < 0$ for all $v\neq 0_{\varrho}$ and $(\tau,\rho)\in\tspc\times\rspc$.}
\end{proof}

\sean{
\begin{remark}
    Corollary~\ref{cor: Matrix Inequality} provides a convenient method for validating the sufficient condition of Theorem~\ref{thm: A is GES}.
    In particular, Corollary~\ref{cor: Matrix Inequality} only requires the checking of two matrix inequalities, namely, $\mmat(0_N,0_{M^\ast})<\mathbf{0}$ and $\mmat(T_2,T_4)<\mathbf{0}$, to ensure $\mmat(\tau,\rho)<\mathbf{0}$ holds for all $(\tau,\rho)\in\tspc\times\rspc$.
    Nevertheless, Corollary~\ref{cor: Matrix Inequality} is not constructive in the sense that it does not provide a method for selecting $\sigma$, $T_1^p$, $T_2^p$, $T_3^r$, $T_4^r$, $K_u$, $K_{\eta,p}$, and $K_{\zeta,pr}$ for all $p\in\verts$ and $r\in[M^\ast]$ so that $\mmat(0_N,0_{M^{\ast}})<\mathbf{0}$ and $\mmat(T_2,T_4)<\mathbf{0}$ hold.
    Furthermore, Corollary~\ref{cor: Matrix Inequality} does not examine how the structure of the clustered network affects the set of matrices used to define $P(\tau,\rho)$ that satisfy $\mmat(0_N,0_{M^{\ast}})<\mathbf{0}$ and $\mmat(T_2,T_4)<\mathbf{0}$.
    Since these topics will require an in-depth analysis, we reserve them for future work.
\end{remark}}

\fred{
In Section~\ref{sec: Numerical}, we will leverage Theorem~\ref{thm: A is GES} and Corollary~\ref{cor: Matrix Inequality} in an application and demonstrate that a set of parameters satisfying $\mmat(0_N,0_{M^{\ast}})<\mathbf{0}$ and $\mmat(T_2,T_4)<\mathbf{0}$ exist.}

\section{Robustness} \label{sec: Robustness}
In this section, we briefly discuss the robustness properties of the set $\attr$ for the hybrid system $\hyb$.
\fred{Since $\hyb$ is nominally well-posed and the set $\attr$ is compact and GES for $\hyb$, standard arguments in~\cite[Chapter 7]{Goebel.Sanfelice.ea2012} and~\cite[Section 3.5.1]{Phillips.Sanfelice2019} can be used to show that $\attr$ is robust to vanishing perturbations for $\hyb$.}
For the remainder of this section, we examine the robustness of $\attr$ to non-vanishing measurement noise\fred{, as defined in Definition~\ref{def: ISS},} injected via a modified jump map of $\hyb$.

The evolution of the variable $\eta_p$ under the hybrid dynamics in~\eqref{eqn: eta_p dynamics} is ideal because the output information in $\{y_q\}_{q\in\nbd_p\cap\cluster_p}$ is measured exactly and without corrupting noise when determined by the timer $\tau_p$.
An analogous comment follows for $\zeta_{pr}$.
Nevertheless, in practice, none-vanishing measurement noise will be present, and its effect on the GES result in Theorem~\ref{thm: A is GES} should be explored to determine the utility of the proposed consensus strategy.
In light of this need, consider the following modifications of~\eqref{eqn: eta_p dynamics} and~\eqref{eqn: zeta_pr dynamics}:
\begin{equation} \label{eqn: noisy eta_p dynamics}
\begin{aligned}
	\dot{\eta}_p &= K_{\eta,p}\eta_p, & \tau_p &\in[0,T_2^p]\\
	\eta_p^+ &= \sum_{q\in\nbd_p\cap\cluster_p}(y_q - y_p) + \delta_p, & \tau_p &= 0
\end{aligned}
\end{equation}
and
\begin{equation} \label{eqn: noisy zeta_pr dynamics}
\begin{aligned}
	\dot{\zeta}_{pr} &= K_{\zeta,pr}\zeta_{pr}, & \rho_r &\in[0,T_4^r]\\
	\zeta_{pr}^+ &= \sum_{q\in\nbd_p\cap\verts^r}(y_q - y_p) + \delta_{pr}, & \rho_r &=0,
\end{aligned}
\end{equation}
where $\delta_p,\delta_{pr}\in\RR^m$ model non-vanishing measurement noise.
Typically, measurement noise is additively injected, e.g., $y_p + \delta_p$, which would imply $\eta_p^+ = \sum_{q\in\nbd_p\cap\cluster_p}(y_q - y_p + \delta_q - \delta_p)$.  
Nevertheless, \fred{with a slight abuse of notation,} we consider the lumped perturbation $\delta_p$ instead of $\sum_{q\in\nbd_p\cap\cluster_p}(\delta_q - \delta_p)$ for convenience.
A similar statement follows for $\delta_{pr}$.
Nearly all other components of $\hyb$, that is, the flow equations of $x^\circ$, $\tilde{\eta}$, $\tilde{\zeta}$, $\tau$, and $\rho$, the definitions in~\eqref{eqn: etaTilde_p} and~\eqref{eqn: zetaTilde_pr}, and the jump equations/inclusions of $x^\circ$, $\tau$, and $\rho$, will remain unchanged.
Yet, under the jump equations of~\eqref{eqn: noisy eta_p dynamics} and~\eqref{eqn: noisy zeta_pr dynamics}, the jump equations of $\tilde{\eta}_p$ and $\tilde{\zeta}_{pr}$ do change.

The definition in~\eqref{eqn: etaTilde_p} and jump equation in~\eqref{eqn: noisy eta_p dynamics} imply $\tilde{\eta}_p^+=\delta_p$ whenever a jump occurs in response to $\tau_p = 0$, and $\tilde{\eta}_p^+=\tilde{\eta}_p$ otherwise.
Likewise, \eqref{eqn: zetaTilde_pr} and the jump equation in~\eqref{eqn: noisy zeta_pr dynamics} imply $\tilde{\zeta}_{pr}^+=\delta_{pr}$ for all $p\in\verts^r$ whenever a jump occurs in response to $\rho_r = 0$, and $\tilde{\zeta}_{pr}^+=\tilde{\zeta}_{pr}$ for all $p\in\verts^r$ otherwise.
With these observations in place, we construct a modification of $\hyb$.

Let $\distspc\subseteq \RR^{mN}\times\RR^{mNM^\ast}$, and consider the hybrid system $\kyb$ with flow set $\widetilde{C}\coloneqq C\times\distspc$, jump set $\widetilde{D}\coloneqq D\times\distspc$, single-valued flow map $\tilde{f}\colon \statespc\times\distspc\to\statespc$ by $\tilde{f}(\xi,\delta) \coloneqq f(\xi)$, and set-valued jump map $\widetilde{G}\colon\statespc\times\distspc\rightrightarrows\statespc$.
Furthermore, let $\delta^r\coloneqq (\delta_{pr})_{p\in\verts}$, $\delta\coloneqq ((\delta_p)_{p\in\verts}, (\delta^r)_{r\in[M^\ast]})\in\mathcal{D}$,
\begin{equation} \label{eqn: noisy jump map}
\begin{aligned}
    \widetilde{G}(\xi,\delta) &\coloneqq \{\widetilde{G}_{\tau,p}(\xi,\delta)\colon (\xi,\delta)\in D_{\tau,p}\times\distspc \text{ for }p\in\verts\} \cup \{\widetilde{G}_{\rho,r}(\xi,\delta)\colon (\xi,\delta)\in D_{\rho,r}\times\distspc\text{ for }r\in[M^\ast]\}, \\
    \widetilde{G}_{\tau,p}(\xi,\delta) &\coloneqq 
    \left[ \begin{array}{c}
    x^\circ \\
    \left[\tilde{\eta}_1^\top,...,\tilde{\eta}_{p-1}^\top,\delta_p^\top,\tilde{\eta}_{p+1}^\top,...,\tilde{\eta}_N^\top\right]^\top \\
    \tilde{\zeta} \\
    \left[\tau_1,...,\tau_{p-1},[T_1^p,T_2^p],\tau_{p+1},...,\tau_N\right]^\top \\
    \rho
    \end{array} \right], \quad \widetilde{G}_{\rho,r}(\xi,\delta) \coloneqq 
    \left[ \begin{array}{c}
    x^\circ \\
    \tilde{\eta} \\
    \left[\tilde{\zeta}_1^\top,...,\tilde{\zeta}_{r-1}^\top,\big(\tilde{\zeta}_{pr}^+\big)_{p\in\verts}^\top,\tilde{\zeta}_{r+1}^\top,...,\tilde{\zeta}_{M^\ast}^\top\right]^\top \\
    \tau \\
    \left[\rho_1,...,\rho_{r-1},[T_3^r,T_4^r],\rho_{r+1},...,\rho_{M^\ast}\right]^\top
    \end{array} \right], \\
    \tilde{\zeta}_{pr}^+ &= 
    \left\{
        \begin{aligned}
            & \delta_{pr}, & p\in\verts^r\phantom{.} \\
            & 0_m, & p\notin\verts^r.
        \end{aligned}
    \right.
\end{aligned}
\end{equation}

\begin{theorem}
    If there exist timer parameters $0<T_1^p\leq T_2^p$ for every $p\in\verts$ and $0< T_3^r\leq T_4^r$ for every $r\in[M^\ast]$, a constant $\sigma$, controller matrix $K_u$, estimator matrices $K_{\eta,p}$ for each $p\in\verts$, estimator matrices $K_{\zeta,pr}$ for every $(p,r)\in\verts\times[M^\ast]$, and symmetric, positive definite matrices $P_1$, $P_2(\tau)$, and $P_3(\rho)$ that satisfy the inequality $\mmat(\tau,\rho) < \mathbf{0}$ for all $(\tau,\rho)\in\tspc\times\rspc$, then the set $\attr$ in~\eqref{eqn: attractor} is ISS for the hybrid system $\kyb$ with data $(\widetilde{C},\tilde{f},\widetilde{D},\widetilde{G})$.
    In particular, for every solution pair $(\phi,\delta)$ of $\kyb$,
    \begin{equation} \label{eqn: ISS bound}
        \vert \phi(t,j) \vert_{\attr} \leq \max\{2\kappa\exp(-\alpha(t+j))\vert \phi(0,0)\vert, 2\kappa_2\vert\delta\vert_{\infty}\},
    \end{equation}
    where $N^\ast\coloneqq N+M^\ast\in\ZZ_{>0}$, $p_{\max} \coloneqq\lambda_{\max}(\diag{P_2(T_2),P_3(T_4)})\in\RR_{>0}$, and\footnote{The matrices $P_2(\tau)$ and $P_3(\rho)$ are defined in~\eqref{eqn: P(tau,rho)}, while the definitions of the constant vectors $T_2$ and $T_4$ are located above~\eqref{eqn: P(tau,rho)}.}
    \begin{equation*}
        \kappa_2 \coloneqq  \sqrt{\frac{p_{\max} N^\ast(2 - \exp(-\mu T_{\min}/\alpha_2))}{\alpha_1(1 - \exp(-\mu T_{\min}/\alpha_2))}}\in\RR_{>0}. 
    \end{equation*}
    Recall, $\kappa$ and $\alpha$ are defined in Theorem~\ref{thm: A is GES}.
    \hfill$\triangle$
\end{theorem}
\begin{proof}
Consider the Lyapunov function candidate in~\eqref{eqn: V} and the hybrid system $\kyb$.
When $(\xi,\delta)\in\widetilde{C}$, the change in $V(\xi)$ is captured by $\dot{V}(\xi)=z^\top \mmat(\tau,\rho) z$.
Since $\mmat(\tau,\rho)$ is negative definite for all $(\tau,\rho)\in\tspc\times\rspc$, the time derivative of~\eqref{eqn: V} can be bounded as in~\eqref{eqn: VDot 2} using the same argument as in the proof of Theorem~\ref{thm: A is GES}.

When $(\xi,\delta)\in\widetilde{D}$, the change in $V(\xi)$
is given by $\Delta V(\xi) = V(g)-V(\xi)$ where $g\in\widetilde{G}(\xi,\delta)$.
Suppose $\tau_q=0$ and $\rho_s=0$ for some $q\in\verts$ and $s\in[M^{\ast}]$.
It then follows that 
\begin{align}
    \Delta V(\xi) &= (z^+)^\top P(\tau^+,\rho^+)(z^+) - z^\top P(\tau,\rho) z \label{eqn: V2 Jump} \\
    &= \delta_q^\top P_{2,q}\exp(\sigma\tau_q^+)\delta_q + (\delta^s)^\top P_{3,s}\exp(\sigma\rho_s^+)\delta^s -\tilde{\eta}_q^\top P_{2,q}\tilde{\eta}_q -\tilde{\zeta}_s^\top P_{3,s}\tilde{\zeta}_s \nonumber \\
    &\leq \delta_q^\top P_{2,q}\exp(\sigma T_2^q)\delta_q + (\delta^s)^\top P_{3,s}\exp(\sigma T_4^s)\delta^s. \nonumber
\end{align}
From~\eqref{eqn: V2 Jump}, we can see that the change in $V(\xi)$ may be positive after a jump, which would cause the value of $V$ to increase.
In the worst-case scenario, i.e., when all agents and inter-clusters experience a simultaneous jump, one has that
\begin{equation} \label{eqn: V2 Jump 2}
    \Delta V(\xi) \leq ((\delta_p)_{p\in\verts})^\top P_2(T_2)(\delta_p)_{p\in\verts} + ((\delta^r)_{r\in[M^\ast]})^\top P_3(T_4)(\delta^r)_{r\in[M^\ast]} \leq p_{\max}\vert \delta \vert_\infty^2.
\end{equation}
Substituting $\Delta V(\xi) = V(g)-V(\xi)$ into~\eqref{eqn: V2 Jump 2} yields
\begin{equation} \label{eqn: V2 Jump Inequality}
    V(g)\leq p_{\max}\vert \delta \vert_\infty^2 + V(\xi).
\end{equation}
Next, fix a solution pair $(\phi,\delta)$ for the hybrid system $\kyb$, select $(t,j)\in\text{dom}(\phi,\delta)$, and let $0= t_0 \leq t_1 \leq ... \leq t_j\leq t$ satisfy
\begin{equation*}
    \text{dom}(\phi,\delta)\bigcap([0,t_j]\times\{ 0,1,...,j-1\}) = \bigcup_{k=1}^j ([t_{k-1},t_k]\times\{k-1\}).
\end{equation*}
Using a similar argument to that in the proof of Theorem~\ref{thm: A is GES} with $(\phi(h,k-1),\delta(h,k-1))\in \widetilde{C}$ for almost all $h\in[t_{k-1},t_k]$ and each $k\in\{1,2,...,j\}$, one can show that
\begin{equation} \label{eqn: V2 solution over flow interval}
    V(\phi(t_k,k-1)) \leq \exp\left(-\frac{\mu}{\alpha_2}(t_k-t_{k-1})\right) V(\phi(t_{k-1},k-1)).
\end{equation}
For each $(\phi(t_k,k-1),\delta((t_k,k-1)))\in \widetilde{D}$ and $k\in\{1,2,...,j\}$, the inequality in~\eqref{eqn: V2 Jump Inequality} implies
\begin{equation} \label{eqn: V2 solution during jumps}
    V(\phi(t_k,k))\leq p_{\max}\vert \delta \vert_\infty^2 + V(\phi(t_k,k-1)).
\end{equation}
One can then inductively stitch~\eqref{eqn: V2 solution over flow interval} and~\eqref{eqn: V2 solution during jumps} along the solution $\phi$ to obtain
\begin{equation} \label{eqn: V2 GUUB Bound 1}
    V(\phi(t,j)) \leq \exp\left(-\frac{\mu}{\alpha_2}t\right) V(\phi(0,0)) + p_{\max}\vert\delta\vert_{\infty}^2 \sum_{s=1}^j\exp\left(-\frac{\mu}{\alpha_2}(t_j - t_s)\right),
\end{equation}
where we used the fact that $t_j \leq t$ and $V(\xi)$ is non-increasing during flows.
A bound for the series in~\eqref{eqn: V2 GUUB Bound 1} is derived next.

Let $j^\ast \coloneqq \lfloor j/N^\ast\rfloor\in\ZZ_{\geq 0}$.
Note $j^\ast N^\ast\leq j$ and $j - j^\ast N^\ast\leq N^\ast$.
The summation in~\eqref{eqn: V2 GUUB Bound 1} can be rewritten as
\begin{equation} \label{eqn: series equality}
    \sum_{s=1}^j\exp\left(-\frac{\mu}{\alpha_2}(t_j - t_s)\right)  =  \sum_{s=0}^{j^\ast - 1} \sum_{k=1}^{N^\ast}\exp\left(-\frac{\mu}{\alpha_2}(t_j - t_{sN^\ast + k})\right) + \sum_{s= j^\ast N^\ast + 1}^j\exp\left(-\frac{\mu}{\alpha_2}(t_j - t_s)\right). 
\end{equation}
Note that $\exp(-\tfrac{\mu}{\alpha_2}(t_j -t_{sN^{\ast}+k})) \leq \exp(-\tfrac{\mu}{\alpha_2}(t_j -t_{(s+1)N^{\ast}}))$ for each $k\in[N^\ast]$ since $t_{sN^{\ast}+1}\leq t_{sN^{\ast}+2}\leq...\leq t_{(s+1)N^{\ast}}\leq t_{j}$.
Therefore,
\begin{equation} \label{eqn: series inequality 1}
    \sum_{k=1}^{N^\ast}\exp\left(-\frac{\mu}{\alpha_2}(t_j - t_{sN^\ast + k})\right)  \leq 
    N^\ast\exp\left(-\frac{\mu}{\alpha_2}t_j^{s+1}\right),
\end{equation}
where $t_j^{s+1}\coloneqq t_j - t_{(s+1)N^\ast}$.
Furthermore, since $\exp(-\tfrac{\mu}{\alpha_2}(t_j - t_s))\leq 1$ provided $t_j - t_s\geq 0$ and $t_{j^\ast N^\ast + 1}\leq t_{j^\ast N^\ast + 2}\leq ... \leq t_j$, one has that
\begin{equation} \label{eqn: series inequality 2}
    \sum_{s= j^\ast N^\ast + 1}^j\exp\left(-\frac{\mu}{\alpha_2}(t_j - t_s)\right) \leq j-j^{\ast}N^{\ast} \leq N^\ast.
\end{equation}
Given~\eqref{eqn: series inequality 1} and~\eqref{eqn: series inequality 2}, the series in~\eqref{eqn: series equality} can be bounded as
\begin{equation} \label{eqn: series inequality 3}
    \sum_{s=1}^j\exp\left(-\frac{\mu}{\alpha_2}(t_j - t_s)\right) \leq \sum_{s=0}^{j^\ast - 1} N^\ast\exp\left(-\frac{\mu}{\alpha_2}t_j^{s+1}\right) + N^\ast. 
\end{equation}
With respect to the right-hand side of~\eqref{eqn: series equality}, the outer summation in the left term describes jumps that occur simultaneously between all agents and inter-clusters; the right term describes individual jumps taken by agents or inter-clusters.
In the case of simultaneous jumps, one can show that $T_{\min}\leq t_{(s+1)N^\ast}-t_{sN^\ast}\leq T_{\max}$ for $s=0,1,...,j^\ast - 1$.
In addition, one can show $t_{j^\ast N^\ast} -t_{(s+1)N^\ast} \geq (j^\ast - (s+1))T_{\min}$ for $s=0,1,...,j^\ast - 1$.
Leveraging $t_j \geq t_{j^\ast N^\ast}$, $t_{j^\ast N^\ast} -t_{(s+1)N^\ast} \geq (j^\ast - (s+1))T_{\min}$, and~\eqref{eqn: series inequality 3}, it follows that
\begin{equation} \label{eqn: series inequality 4}
    \sum_{s=1}^j\exp\left(-\frac{\mu}{\alpha_2}(t_j - t_s)\right) \leq \sum_{s=0}^{j^\ast - 1} N^\ast\exp\left(-s\frac{\mu}{\alpha_2} T_{\min}\right) + N^\ast. 
\end{equation}
Note $j^\ast \to\infty$ as $j\to\infty$ since $j^\ast = \lfloor j/N^\ast\rfloor$.
Moreover, since $\exp(-\mu T_{\min}/\alpha_2)\in (0,1)$, the corresponding geometric series converges, that is,
\begin{equation} \label{eqn: geometric series}
    \sum_{s=0}^{\infty} N^\ast\exp\left(-s\frac{\mu}{\alpha_2} T_{\min}\right) = N^\ast\sum_{s=0}^{\infty} \exp\left(-\frac{\mu}{\alpha_2} T_{\min}\right)^s = \frac{N^\ast}{1 - \exp(-\mu T_{\min}/\alpha_2)}.
\end{equation}
Thus, \eqref{eqn: series inequality 4} and~\eqref{eqn: geometric series} imply
\begin{equation} \label{eqn: series inequality 5}
    \sum_{s=1}^j\exp\left(-\frac{\mu}{\alpha_2}(t_j - t_s)\right) \leq \frac{N^\ast(2 - \exp(-\mu T_{\min}/\alpha_2))}{1 - \exp(-\mu T_{\min}/\alpha_2)}. 
\end{equation}
The substitution of~\eqref{eqn: series inequality 5} into~\eqref{eqn: V2 GUUB Bound 1} yields
\begin{equation} \label{eqn: V2 GUUB Bound 2}
    V(\phi(t,j)) \leq \exp\left(-\frac{\mu}{\alpha_2}t\right) V(\phi(0,0)) + p_{\max}\vert\delta\vert_{\infty}^2 N^\ast \frac{2 - \exp(-\mu T_{\min}/\alpha_2)}{1 - \exp(-\mu T_{\min}/\alpha_2)}.
\end{equation}
Upon substituting~\eqref{eqn: t+j inequality} into~\eqref{eqn: V2 GUUB Bound 2} and using the bounds in~\eqref{eqn: V Bounds}, the desired result in~\eqref{eqn: ISS bound} follows.
\end{proof}

\section{Simulation Example \& Discussion} \label{sec: Numerical}
To confirm the theoretical results, we consider a rendezvous application for a MAS of identical marine crafts maneuvering on a common plane.
The kinematics of agent $p$ are
\begin{equation} \label{eqn: marine craft kinematics}
    \dot{c}_{1,p} = v_p\cos(\theta_p), \ \ \dot{c}_{2,p} = v_p\sin(\theta_p), \ \ \dot{\theta}_p = \omega_p.
\end{equation}
For each $p\in\verts$, \fred{suppose the center of mass (CM) of agent $p$ is denoted by $c_p\coloneqq (c_{1,p},c_{2,p})\in\RR^2$, expressed in a common inertial reference frame, and fixed relative to the body of the craft.}
The heading and angular velocity are denoted by $(\theta_p,\omega_p)\in (-\pi,\pi]\times\RR$, and the linear speed is denoted by $v_p\in\RR$.
The control input of agent $p$ is $\mu_p\coloneqq (v_p,\omega_p)\in\RR^2$.

\begin{figure}[t]
\centering 
\includegraphics[width=0.49\columnwidth]{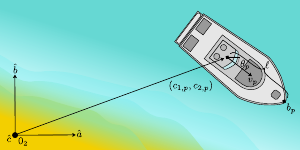} 
\caption{A common inertial reference frame is depicted on the bottom left with origin $0_2$ and right-handed coordinate system defined by the orthogonal unit vectors $\hat{a}$, $\hat{b}$, and $\hat{c}$.
The positions of the CM and bow of agent $p$ are given by $c_p$ and $b_p$, respectively.
The distance between the CM and bow is denoted by $\ell$.
The speed along the unit vector $(b_p - c_p)/\Vert b_p - c_p \Vert$ and heading relative of $\hat{a}$ of agent $p$ are denoted by $v_p$ and $\theta_p$, respectively.
}
\label{fig: Boat}
\end{figure}

For each marine craft $p\in\verts$, let $\ell\in\RR_{>0}$ denote the length between the CM and bow.
Using the CM position $c_p$, heading $\theta_p$, and length $\ell$, the position of the bow of agent $p$ is given by $b_p\coloneqq c_p + \ell(\cos(\theta_p),\sin(\theta_p))$.
Figure~\ref{fig: Boat} provides a clarifying illustration.
The goal of this simulation example is to achieve rendezvous in the bow positions, that is, $b_p=b_q$ for all $p,q\in\verts$.
To enable the use of our consensus result, we introduce a transformation.
Utilizing~\eqref{eqn: marine craft kinematics}, the time derivative of $b_p$ is
\begin{equation} \label{eqn: transformation}
    \dot{b}_p = 
    \underbrace{\begin{bmatrix}
        \cos(\theta_p) & -\sin(\theta_p) \\
        \sin(\theta_p) & \phantom{-}\cos(\theta_p)
    \end{bmatrix}}_{\coloneqq R(\theta_p)}
    \underbrace{\begin{bmatrix}
        1 & 0 \\
        0 & \ell 
    \end{bmatrix}}_{\coloneqq D}\mu_p.
\end{equation}
Since $R(\theta_p)$ is a rotation matrix and $\ell>0$, $R(\theta_p)^{-1}$ and $D^{-1}$ are always well-defined.
Furthermore, for any velocity $\dot{b}_p$, the corresponding control input is given by $\mu_p = D^{-1}R(\theta_p)^{-1}\dot{b}_p$.
Given a continuously differentiable reference trajectory for the bow position $b_p$, one can compute the corresponding control input for~\eqref{eqn: marine craft kinematics}, pointwise in time, by utilizing the inverse map $\linop^{-1}(\theta_p,s)\coloneqq D^{-1}R(\theta_p)^{-1}s$.
Next, we provide a method for producing a reference trajectory for each $b_p$.

For each agent $p\in\verts$, consider the auxiliary system
\begin{equation} \label{eqn: virtual system of agent p}
    \mtt\ddot{z}_p + \ctt\dot{z}_p = u_p,
\end{equation}
where $z_p\in\RR^2$ denotes a virtual position, $u_p\in\RR^2$ is a virtual control input, $\mtt\in\RR^{2\times2}$ is a positive definite mass matrix, and $\ctt\in\RR^{2\times 2}$ is a non-zero viscous damping matrix.
The second-order ODE in~\eqref{eqn: virtual system of agent p} can be expressed as an LTI system by using \fred{$x_{1,p}\coloneqq z_p\in\RR^2$, $x_{2,p}\coloneqq \dot{z}_p\in\RR^2$, $x_p\coloneqq (x_{1,p},x_{2,p})\in\RR^4$ using the notation from Section~\ref{sec: Notation}, and
\begin{equation} \label{eqn: LTI matrices}
    A \coloneqq \begin{bmatrix}
    0_{2\times 2} & I_2 \\
    0_{2\times 2} & -\mtt^{-1}\ctt
    \end{bmatrix}\in\RR^{4\times 4}, \ 
    B \coloneqq \begin{bmatrix}
        0_{2\times 2} \\
        \mtt^{-1}
    \end{bmatrix}\in\RR^{4\times 2}, \
    H \coloneqq \begin{bmatrix}
         I_2 & 0_{2\times 2}
    \end{bmatrix}\in\RR^{2\times 4}.
\end{equation}}
Note $(A,B)$ is controllable, $(A,H)$ is observable, and position is assumed measurable given the definition of $H$.

The ODE in~\eqref{eqn: virtual system of agent p} complies with the LTI model requirement of our consensus strategy and can model the motion of a point-mass under the influence of viscous damping.
Furthermore, the ODE in~\eqref{eqn: virtual system of agent p} will be used to generate an almost-everywhere continuously differentiable reference trajectory for the bow of agent $p$, which will be tracked using $\linop^{-1}(\theta_p,x_{2,p})$.
We now describe an implementation of this reference trajectory tracking strategy.
Since $x_{1,q}$ is the virtual position variable of agent $q$, which is not a physical quantity that can be independently measured by agent $p\neq q$, the controller comprised of~\eqref{eqn: tau timer p}--\eqref{eqn: agent controller} is not directly implementable.
Nevertheless, knowledge of $x_{1,q}$ can be granted via wireless communication and computation.
For example, consider agent $p$. 
Whenever agent $p$ updates $\eta_p$ or $\zeta_{pr}$ for some $r\in [M^\ast]$, agent $p$ can broadcast the updated quantity to all neighboring agents $q\in\nbd_p$.
If $K_{\eta,p} = K_{\eta,q}$ and $K_{\zeta,pr} = K_{\zeta,qs}$ for all $p,q\in\verts$ and $r,s\in[M^\ast]$ or if these matrices are distinct and known \textit{a priori} by all agents, then $\eta_p$ and $\zeta_{pr}$ can be computed independently and simultaneously by all neighbors of agent $p$ given the flow dynamics in~\eqref{eqn: eta_p dynamics} and~\eqref{eqn: zeta_pr dynamics}.
Hence, each neighbor of agent $p$ can compute the control input of agent $p$, where exact model knowledge of~\eqref{eqn: virtual system of agent p} enables each neighbor of agent $p$ to compute $x_{1,p}$.
By repeating this strategy for every neighbor $k\in\nbd_q$, agent $q$ can implement~\eqref{eqn: tau timer p}--\eqref{eqn: agent controller} to obtain a reference trajectory for itself.

By driving the auxiliary variables in $\{x_p\}_{p\in\verts}$ into consensus, which will be achieved using the result from Section~\ref{sec: Modeling} and applying the transformation $\linop^{-1}$ pointwise in continuous-time for each agent $p\in\verts$, the marine crafts can accomplish rendezvous with respect to the bow positions.
\fred{Note, however, our result does not guarantee collision avoidance between the marine crafts.
Therefore, the trajectories generated by this approach can be treated as references to be tracked by a controller capable of ensuring collision avoidance.}
Two simulation cases are considered, namely, a nominal simulation that is free of perturbations and a more realistic simulation that contains measurement noise.
The simulation parameters are \fred{the following:} $N=14$, $n = 4$, $d = m = 2$, $M = 5$, $M^\ast = 8$, $\ell = 0.53$, $\sigma = 500$, $\mtt = I_2$, $\ctt = 2.66\cdot I_2$, $K_u = 0.5\cdot I_2$, and, for all $p\in\verts$ and $r\in[M^\ast]$, $T_1^p = T_3^r = 0.001$, $T_2^p = T_4^r = 0.01$,
\begin{equation*}
    K_{\eta,p} = K_{\zeta,pr} = 
    \begin{bmatrix}
        -2.33 & -1.33 \\
        \phantom{-}1.33 & -5.66
    \end{bmatrix}.
\end{equation*}
Moreover, $x_{1,p}(0,0)=b_p(0,0)$ for each $p\in\verts$.
\fred{Thus, there are $14$ agents in total that are divided into $5$ clusters and $8$ inter-clusters.
The parameters $n$, $d$, and $m$ are determined by the LTI system corresponding to the auxiliary system in~\eqref{eqn: virtual system of agent p}.
The length $\ell$ between the bow and CM of each marine craft was chosen arbitrarily.
Likewise, the parameters $\sigma$, $T_1^p$, $T_3^r$, $T_2^p$, and $T_4^r$ and matrices $\mtt$, $\ctt$, $K_u$, $K_{\eta,p}$, and $K_{\zeta,pr}$ were chose at will provided $0<T_1^p\leq T_2^p$ for each agent $p\in\verts$, $0<T_3^r\leq T_4^r$ for each inter-cluster $r\in[M^\ast]$, and the matrix inequality in~\eqref{eqn: Theorem 1 items} could be satisfied at $(\tau,\rho)=(0_N,0_{M^\ast})$ and $(\tau,\rho)=(T_2,T_4)$.}
Under these parameters, MATLAB and the CVX Toolbox~\cite{Grant.Boyd2008,Grant.Boyd2014} were employed to solve the matrix inequality $\mmat(\tau,\rho) \leq \mathbf{0}$ for points $(\tau,\rho)= (0_N,0_{M^\ast})$ and $(\tau,\rho)= (T_2,T_4)$.
Consequently, Corollary~\ref{cor: Matrix Inequality} implies the sufficient conditions of Theorem~\ref{thm: A is GES} are satisfied.
Unfortunately, we are unable to provide the solution to the matrix inequality since $P(0_N,0_{M^\ast})\in\RR^{304\times 304}$ in our application.
Note, CVX took about 100 seconds to compute $P(0_N,0_{M^\ast})$.

\begin{figure}[t]
\centering 
\includegraphics[width=0.49\columnwidth]{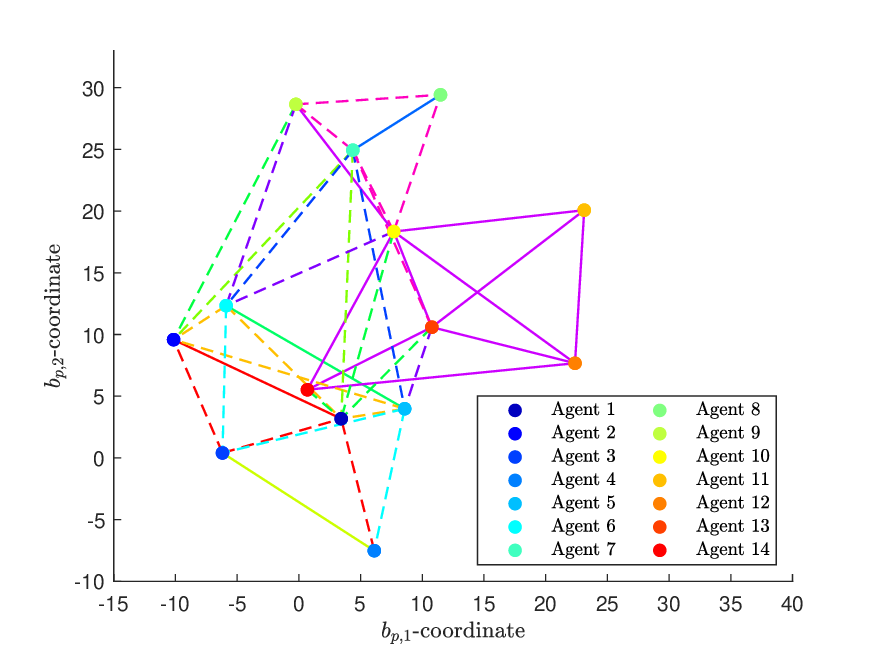} 
\caption{Illustration of the clustered network used in the nominal and perturbed simulation cases.
The nodes and edges of the communication graph $\graph$ are represented by the disks and both solid and dashed lines, respectively.
Different colored disks connected by solid lines of the same color represent a cluster sub-graph.
Different colored disks connected by dashed lines of the same color represent an inter-cluster sub-graph.
The planar position of each node is coincident with the initial bow position of the corresponding agent.
}
\label{fig: Network Topology}
\end{figure}

The parameters and communication topology, i.e., the clustered network depicted in Figure~\ref{fig: Network Topology}, are identical for the nominal and perturbed simulation cases.
\fred{Moreover, the initial condition of each agent was selected in a manner leading to the construction of a distance-dependent graph with nodes in $\RR^4$ having edges no longer than $25$ units of length.}
Figures~\ref{fig: Nominal Trajectories}--\ref{fig: Nominal Quad Plot} correspond to the simulation results for the nominal case, while Figures~\ref{fig: Noisy Trajectories}--\ref{fig: Noisy Quad Plot} correspond to the simulation results for the perturbed case.
To examine the robustness of our consensus controller to various perturbation magnitudes and communication graphs, a modest Monte Carlo simulation was performed with results provided in Figure~\ref{fig: Rollout Trajectories of xCirc}.
Note only the graph $\graph$, perturbation magnitudes, and controller matrix $K_u$ were changed in the Monte Carlo simulation.
\fred{The graphs were defined in a similar fashion to the nominal and perturbation simulation cases described above, leading to corresponding initial conditions for each agent.}
The controller matrix $K_u = 1.5\cdot I_2$ was fixed for all $30$ runs of the Monte Carlo simulation.
\fred{However, perturbation magnitudes were randomly selected from the interval $[0,0.075]$ using a uniform distribution.}
The partition of each graph's node set $\verts$ was also randomly selected.

\subsection{Consensus under Ideal Conditions}
\begin{figure}[t]
\centering 
\includegraphics[width=0.49\columnwidth]{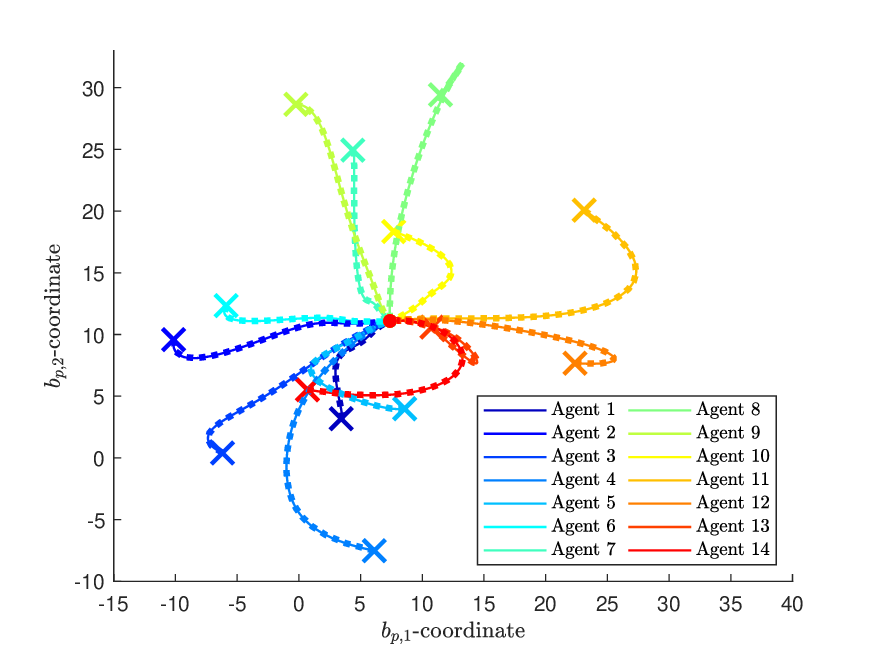} 
\caption{Illustration of the reference bow position trajectories and true bow position trajectories for each agent $p\in\verts$.
Solid lines represent the trajectories of $\{x_{1,p}\}_{p\in\verts}$ (reference bow positions), while dotted lines represent the trajectories of $\{b_p\}_{p\in\verts}$ (true bow positions).
The $\times$'s and $\bullet$'s depict the initial and final conditions of all trajectories, respectively.
Observe rendezvous is achieved in both the reference bow positions $\{x_{1,p}\}_{p\in\verts}$ and true bow positions $\{b_p\}_{p\in\verts}$.
}
\label{fig: Nominal Trajectories}
\end{figure}

\begin{figure}[t]
\centering 
\includegraphics[width=0.49\columnwidth]{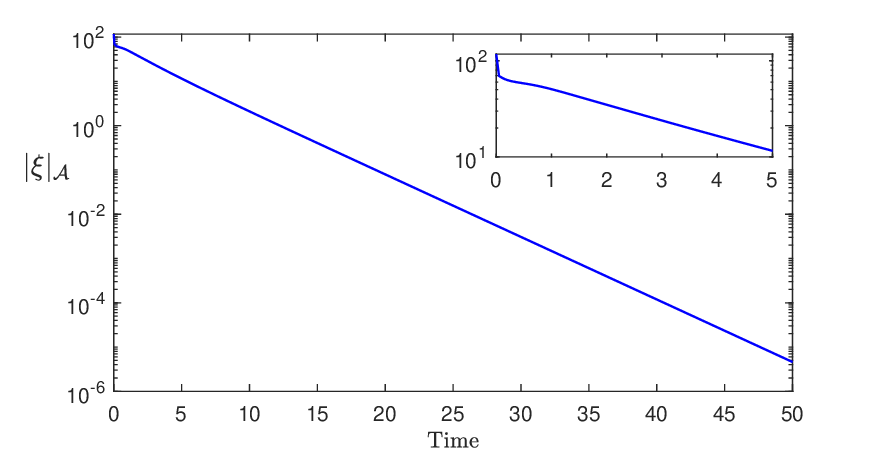} 
\caption{Depiction of $\vert\xi\vert_{\attr}$ as a function of continuous time, where $\xi=\phi(t,j)$---the particular solution to the hybrid system $\hyb$ generated in the nominal simulation.
The vertical axis uses a logarithmic scale; the horizontal axis uses a linear scale.
Based on the plot, the distance of $\xi$ to $\attr$ is exponentially converging to $0$.
}
\label{fig: Nominal Distance to Attractor}
\end{figure}

\begin{figure}[t]
\centering
    \begin{subfigure}[b]{0.49\columnwidth}
         \centering
         \includegraphics[width=1\columnwidth]{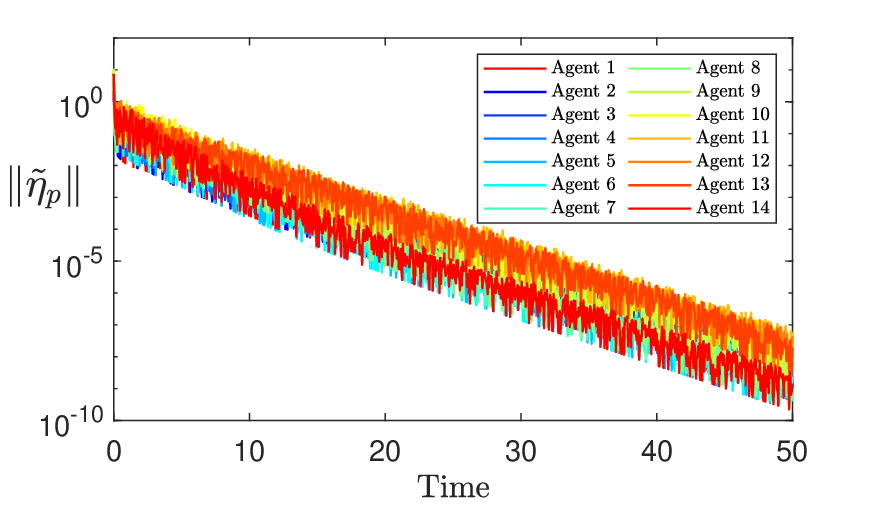}
    \end{subfigure}
    \begin{subfigure}[b]{0.49\columnwidth}
         \centering
         \includegraphics[width=1\columnwidth]{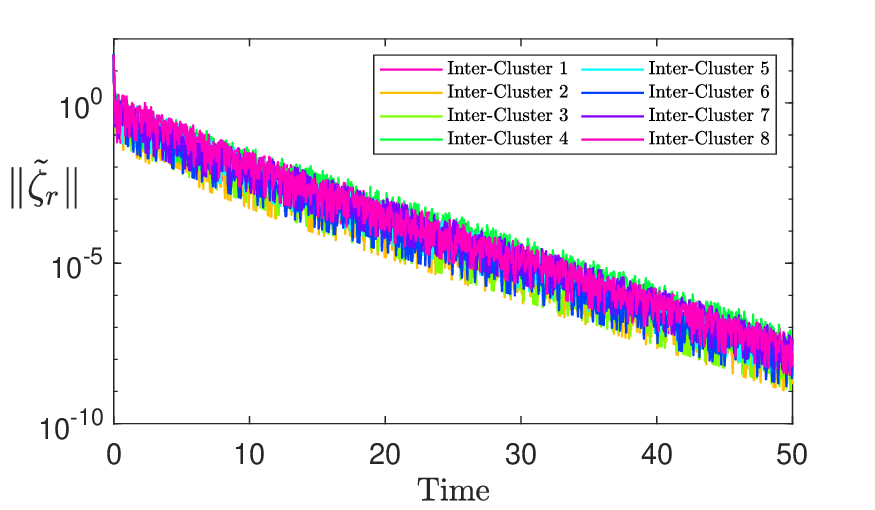} 
    \end{subfigure}
    \begin{subfigure}[b]{0.49\columnwidth}
         \centering
         \includegraphics[width=1\columnwidth]{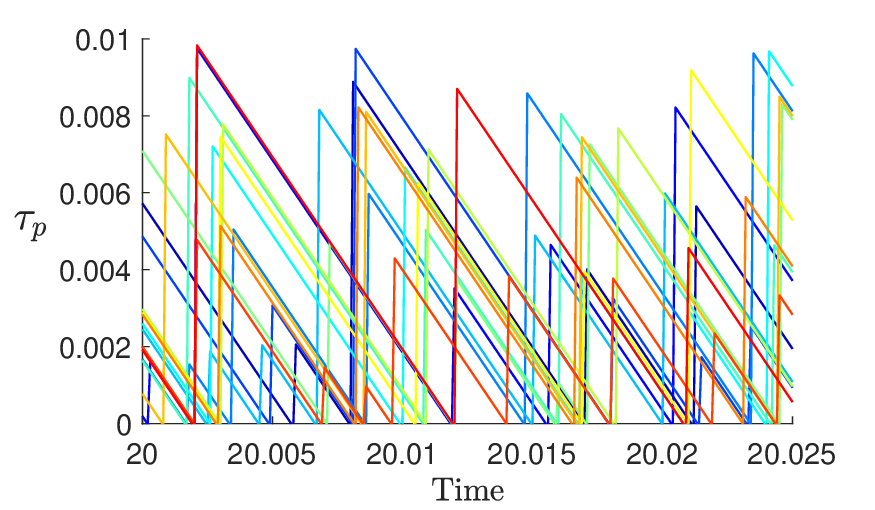}
    \end{subfigure}
    \begin{subfigure}[b]{0.49\columnwidth}
         \centering
         \includegraphics[width=1\columnwidth]{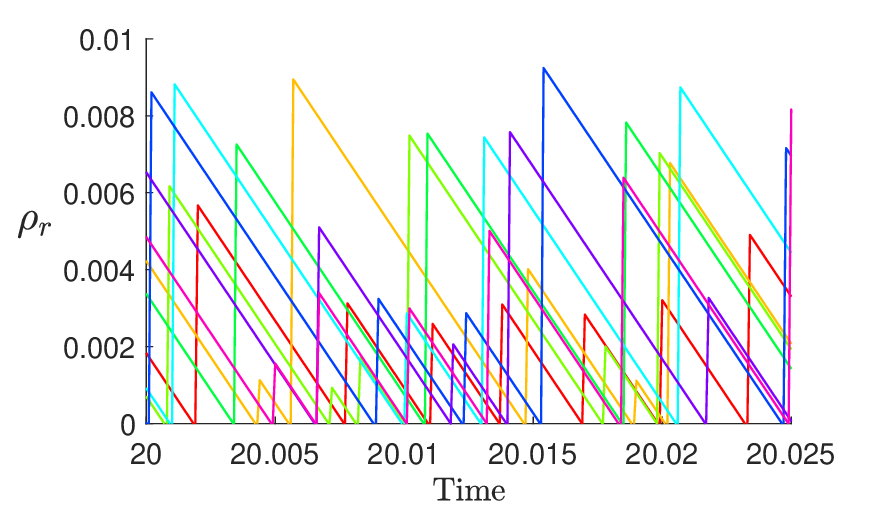} 
    \end{subfigure}
\caption{
\textbf{(Top Left)} Depiction of $\Vert \tilde{\eta}_p(\phi(t,j))\Vert$, for each agent $p\in\verts$, versus continuous time.
Note that the vertical axis uses a logarithmic scale, while the horizontal axis uses a linear scale.
Although noisy, the magnitude of $\tilde{\eta}_p$, along the solution $\phi(t,j)$, tends to $0$ as $t+j$ approaches infinity. 
\textbf{(Bottom Left)} Depiction of the trajectories of $\tau_p(\phi(t,j))$ for each $p\in\verts$.
Since $\tau_p$ determines when agent $p$ updates the $\eta_p$ component of its controller (i.e., when $\tau_p=0$) and the timers in $\{\tau_q\}_{q\in\verts}$ reach $0$ at different times, the variables in $\{\eta_q\}_{q\in\verts}$ are updated intermittently and asynchronously with respect to each other.
\textbf{(Top Right)} Depiction of $\Vert \tilde{\zeta}_r(\phi(t,j))\Vert$, for each $r\in[M^\ast]$, versus continuous time.
Note that the vertical axis uses a logarithmic scale, while the horizontal axis uses a linear scale.
Although noisy, the magnitude of $\tilde{\zeta}_r$, along the solution $\phi(t,j)$, tends to $0$ as $t+j$ approaches infinity.
\textbf{(Bottom Right)} Depiction of the trajectories of $\rho_r(\phi(t,j))$ for each $r\in[M^\ast]$.
Since $\rho_r$ determines when the agents in inter-cluster $r$ update the $\zeta_{qr}$ component of their controller, for $q\in\verts^r$, and the timers in $\{\rho_s\}_{s\in\verts}$ reach $0$ at different times, the variables in $\{\zeta_{qr}\}_{q\in\verts^r}$ are updated intermittently and asynchronously with respect to the variables in $\{\zeta_{qs}\}_{q\in\verts^s}$ for $r\neq s\in [M^\ast]$.
Note that, for each $p\in\verts$, the color of $\Vert\tilde{\eta}_p\Vert$ matches that of $\tau_p$.
Likewise, for each $r\in[M^\ast]$, the color of $\Vert\tilde{\zeta}_r\Vert$ matches that of $\rho_r$.
}
\label{fig: Nominal Quad Plot}
\end{figure}

As validated by Figures~\ref{fig: Nominal Trajectories}--\ref{fig: Nominal Quad Plot}, the proposed control strategy in Section~\ref{sec: Modeling} renders the set $\attr$ \fred{GES} under ideal conditions (i.e., under the absence of measurement noise).
Consequently, the reference trajectories for the marine crafts can be synchronized exponentially fast while using intermittent and asynchronous output feedback (that is, position measurements).
Also, the map $(\theta_p,x_{2,p})\mapsto D^{-1}R(\theta_p)^{-1}x_{2,p}$ enabled each agent $p\in\verts$ to track their reference trajectory.
Recall $x_{2,p}$ denotes the velocity of the reference trajectory for agent $p$.
Therefore, we were able to achieve exponentially fast rendezvous for marine crafts modeled by~\eqref{eqn: marine craft kinematics}.

Of course, driving $\vert\xi\vert_{\attr}$ to $0$ in practice would probably be undesirable as the user may not wish for potentially expensive marine crafts to crash into each other.
Thus, achieving $\vert\xi\vert_{\attr}\leq \omega$ for some appropriate $\omega>0$ may suffice instead.
Given such a scenario, $\vert\xi\vert_{\attr}\leq \omega$, \eqref{eqn: V Bounds}, and~\eqref{eqn: V GES Bound} facilitate the calculation of a conservative, albeit finite, continuous time for achieving $\omega$-approximate rendezvous (see~\cite[Defintion 2]{Zegers.Guralnik.ea2021}).
In particular,
\begin{equation*}
     t^{\ast}\coloneqq\frac{2\alpha_2}{\mu}\ln\left( \sqrt{\frac{\alpha_2}{\alpha_1}}\frac{\vert \phi(0,0) \vert_{\attr}}{\omega}\right) \leq t \text{ and } 0 < \omega < \sqrt{\frac{\alpha_2}{\alpha_1}}\vert\phi(0,0)\vert_{\attr}
     \implies \vert \phi(t,j) \vert_{\attr} \leq \omega
\end{equation*}
for all $(t,j)\in\dom\phi$ such that $(t,j)\geq (t^\ast,j^\ast)$ with $(t^\ast,j^\ast)\in\dom\phi$.
The time $t^{\ast}$ denotes when $\omega$-approximate rendezvous is guaranteed to occur, which depends on the initial condition and control parameters.
The inequality constraint on $\omega$ ensures the logarithm above is positive, yielding a well-defined future time $t^{\ast}$.
While the nominal simulation provides good results, non-vanishing measurement noise will affect a cyber-physical system in practice.
Ergo, the following simulation is motivated.

\subsection{Consensus under Measurement Noise}
To simulate non-vanishing measurement noise, we employ the following: for all $p\in\verts$ and $r\in [M^\ast]$, let
\begin{equation} \label{eqn: perturbations}
    \delta_p = 0.07\sin(0.015t)\sin(0.2t)\cdot 1_m \quad \delta_{pr} = 0.04\sin(0.095t)\cos(0.4t)\cdot 1_m,
\end{equation}
where the perturbations $\delta_p$ and $\delta_{pr}$ are explicit functions of continuous time.
Although all agents are subjected to the same perturbations, the induced affects are different because agents introduce samples of these perturbations are different moments in time, potentially, given the jump map in~\eqref{eqn: noisy jump map}.

\begin{figure}[t]
\centering 
\includegraphics[width=0.49\columnwidth]{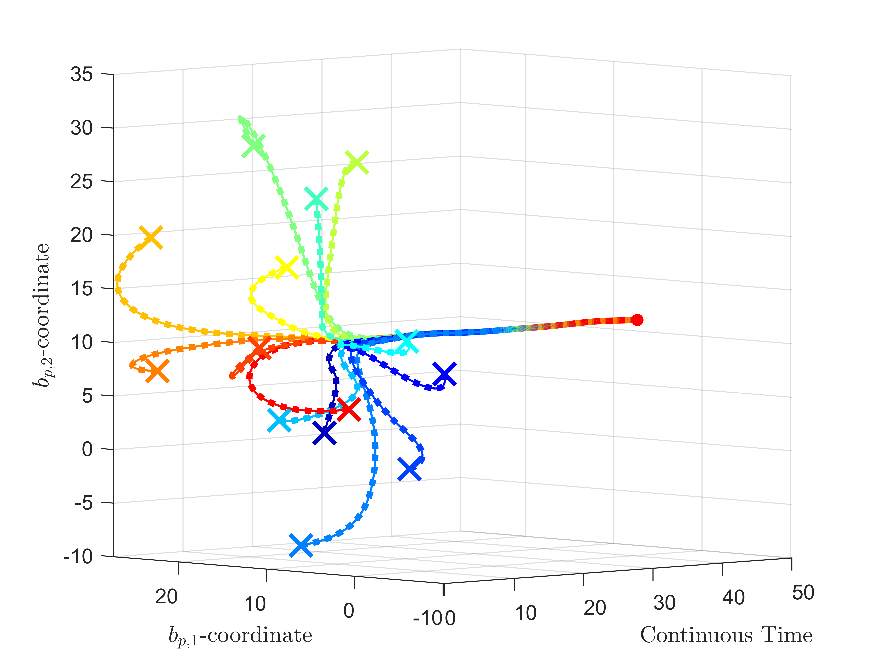} 
\caption{Illustration of the reference bow position trajectories and true bow position trajectories for each agent $p\in\verts$.
Solid lines represent the trajectories of $\{x_{1,p}\}_{p\in\verts}$ (reference bow positions), while dotted lines represent the trajectories of $\{b_p\}_{p\in\verts}$ (true bow positions).
The $\times$'s and $\bullet$'s depict the initial and final conditions of all trajectories, respectively.
Observe rendezvous is achieved in both the reference bow positions $\{x_{1,p}\}_{p\in\verts}$ and true bow positions $\{b_p\}_{p\in\verts}$ despite the presence of measurement noise.
Nevertheless, unlike the nominal simulation, the rendezvous point moves slightly under the influence of measurement noise.}
\label{fig: Noisy Trajectories}
\end{figure}

\begin{figure}[t]
\centering
    \begin{subfigure}[b]{0.49\columnwidth}
         \centering
         \includegraphics[width=1\columnwidth]{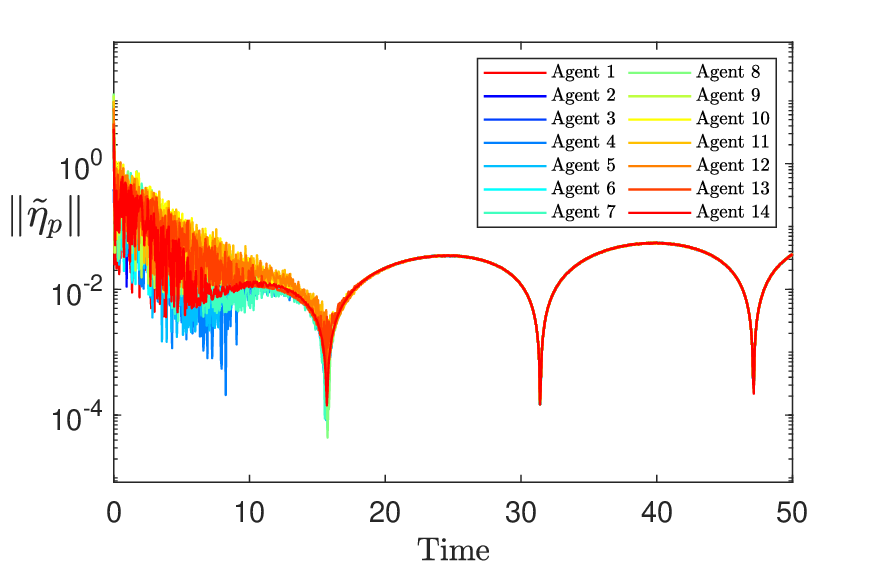}
    \end{subfigure}
    \begin{subfigure}[b]{0.49\columnwidth}
         \centering
         \includegraphics[width=1\columnwidth]{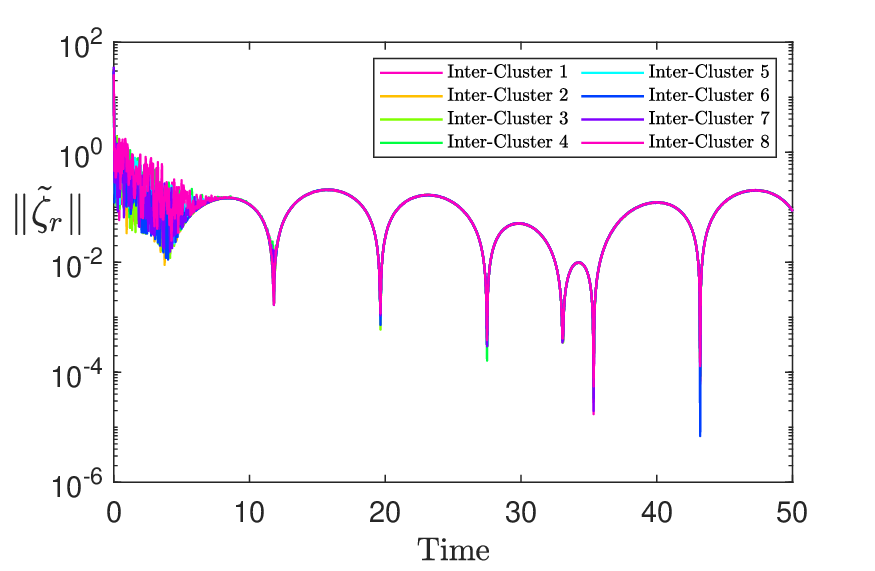} 
    \end{subfigure}
    \begin{subfigure}[b]{0.49\columnwidth}
         \centering
         \includegraphics[width=1\columnwidth]{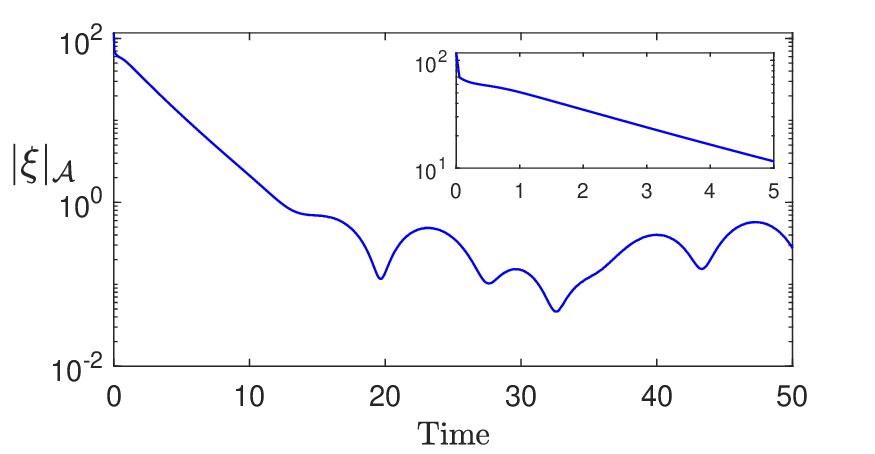}
    \end{subfigure}
    \begin{subfigure}[b]{0.49\columnwidth}
         \centering
         \includegraphics[width=1\columnwidth]{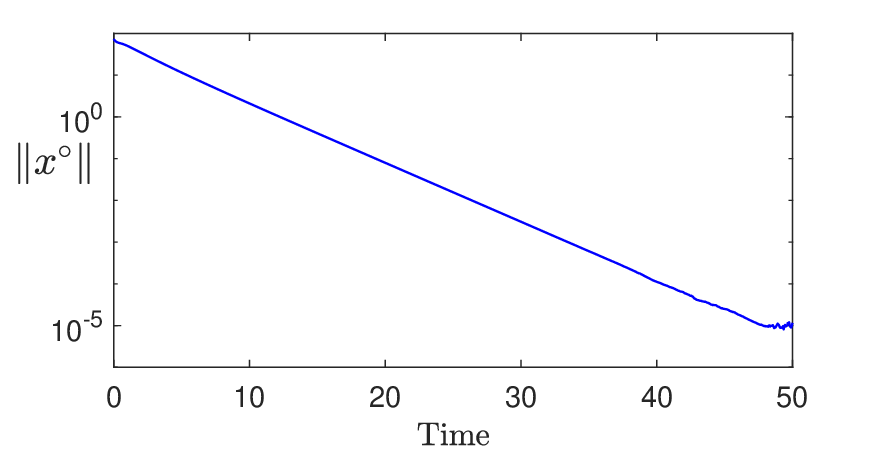} 
    \end{subfigure}
\caption{
\textbf{(Top Left)} Depiction of $\Vert \tilde{\eta}_p(\phi(t,j))\Vert$, for each $p\in\verts$, versus continuous time.
Note the vertical axis uses a logarithmic scale, while the horizontal axis uses a linear scale.
The large oscillations in the trajectories are caused by the sinusoidal measurement noise. 
\textbf{(Bottom Left)} Illustration of the trajectory of $\vert \xi \vert_{\attr}$, where $\xi=\phi(t,j)$---the particular solution to the hybrid system $\kyb$ generated in the simulation.
Observe that the measurement noise causes $\xi$ to converge to an $\omega$-fattening of $\attr$ for some constant $\omega\in\RR_{>0}$.
\textbf{(Top Right)} Depiction of $\Vert \tilde{\zeta}_r(\phi(t,j))\Vert$, for each $r\in[M^\ast]$, versus continuous time.
Observe that the vertical axis uses a logarithmic scale, while the horizontal axis uses a linear scale.
The large oscillations in the trajectory are due to the sinusoidal measurement noise.
\textbf{(Bottom Right)} Illustration of $\Vert x^{\circ}(\phi(t,j))\Vert$ versus continuous time.
Despite the presence of measurement noise, the disagreement in $\{x_p\}_{p\in\verts}$ can be made small and oscillates around a magnitude of $1\times 10^{-5}$.}
\label{fig: Noisy Quad Plot}
\end{figure}

With respect to Figures~\ref{fig: Noisy Trajectories}--\ref{fig: Noisy Quad Plot}, the non-vanishing measurement noise has a clear effect on the trajectories of the MAS.
Minor oscillations in the rendezvous point, $\tilde{\eta}_p$ for all $p\in\verts$, $\tilde{\zeta}_r$ for all $r\in[M^\ast]$, and $\vert\xi\vert_{\attr}$ are apparent.
Nevertheless, $\omega$-approximate rendezvous is accomplished for $\omega\leq 1\times 10^{-4}$.

\begin{figure}[t]
\centering 
\includegraphics[width=0.49\columnwidth]{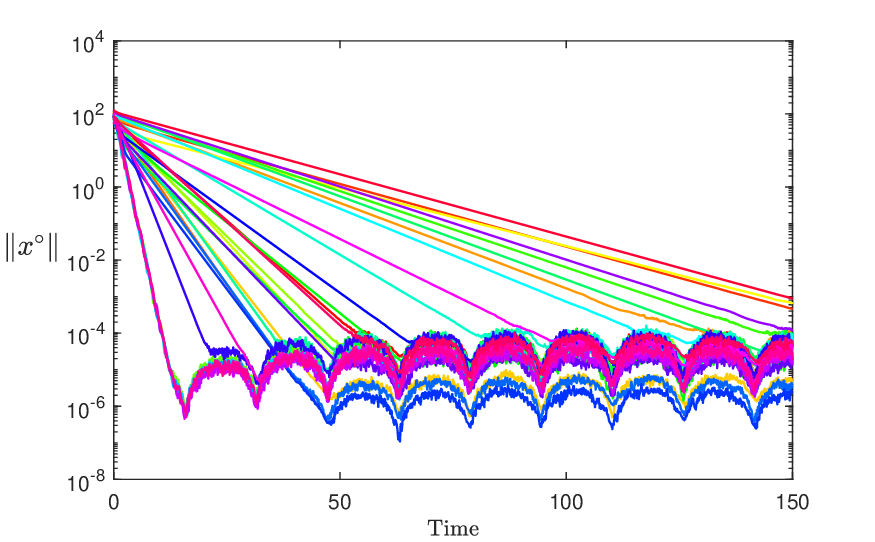} 
\caption{Illustration of $\Vert x^\circ(\phi(t,j)) \Vert$ versus continuous time for $30$ different simulations.
Each color corresponds to a unique scenario and a distinct solution $\phi$ to $\kyb$.
Nearly all simulation parameters are unchanged between runs except for the clustered network and perturbation magnitude.
In all cases, the MAS drove $\Vert x^\circ(\phi(t,j)) \Vert$ to a small neighborhood of $0$.
}
\label{fig: Rollout Trajectories of xCirc}
\end{figure}

To further examine the robustness of the consensus result to measurement noise, $30$ additional simulations using randomly selected clustered networks and perturbation magnitudes were conducted.
Nearly all simulation parameters were unchanged, except $K_u = 1.5\cdot I_2$.
Figure~\ref{fig: Rollout Trajectories of xCirc} conveys the results of these simulations.
In all cases, $\omega$-approximate rendezvous was achieved.

\section{Conclusion}
This paper develops a distributed consensus controller that is hybrid in structure for a clustered network of identical LTI systems utilizing intermittent and asynchronous output feedback.
Specifically, the controller \fred{of each agent} consists of multiple components that are intermittently updated at different rates with output information from \fred{same-cluster neighbors and same-inter-cluster neighbors. 
Although the communication network is modeled as a static, connected, and undirected graph, the flow out output information over the edges of this graph can be intermittent and asynchronous.}

Future work may consider nonlinear, uncertain, and/or heterogeneous agent models \fred{as well as directed communication graphs given their practical importance.
One way of extending the result to heterogeneous agent dynamics is by approximating each dynamics with a common LTI system and then designing the controller of each agent to either adaptively or robustly compensate for the disturbance defined by the difference between the LTI system and the true agent dynamics.}
One may examine consensus under path constraints to guarantee safety or other desired behaviors.
More advanced \fred{estimators} can also be developed to promote less frequent communication and/or measurements.
Of particular importance is the derivation of a constructive extension of Corollary~\ref{cor: Matrix Inequality} to simplify the design of the proposed consensus controller.
Moreover, the effect of the clustered network structure on the solution space of $\mmat(\tau,\rho)<\mathbf{0}$ should be investigated.

\appendix
\section{Proof of Lemma~\ref{lemma: L and S Identities}} \label{appen: L and S Identities}

Since the graph $\graph$ is undirected and defined on $N$ nodes, the associated Laplacian matrix $\lap\in\RR^{N\times N}$ is symmetric, which implies $\lap$ is diagonalizable and has real eigenvalues.
Further, since $\graph$ is static, the diagonalization of $\lap$ is fixed.
Let $\gamma \coloneqq \{w_1,w_2,...,w_N\}\subset\RR^N$ and $\{\lambda_1,\lambda_2,...,\lambda_N\}\subset\RR$ denote the eigenvector and eigenvalue sets of $\lap$, respectively, where $\lambda_p$ is the eigenvalue corresponding to eigenvector $w_p$ for each $p\in[N]$.
Since $\graph$ is connected, $w_1\in\Span{1_N}$, $\lambda_1 = 0$, and, without loss of generality, $\lambda_1<\lambda_2\leq ... \leq \lambda_N$.
Observe that $\gamma$ is an ordered basis for $\Range{\lap}$.
We can then employ the Gram-Schmidt process on $\gamma$ to construct an orthogonal basis $\beta^\prime=\{v_1^\prime, v_2^\prime,...,v_N^\prime\}$ for $\Range{\lap}$, where the normalization of the elements in $\beta^\prime$ yields an orthonormal basis $\beta$.
That is, $\beta=\{v_1, v_2,..., v_N\}\subset\RR^N$, where $v_p \coloneqq v_p^\prime/\Vert v_p^\prime \Vert$ and $v_p\perp v_q$ for all distinct $p,q \in [N]$.
Observe $\beta^\prime$ and $\beta$ are not necessarily unique.
Since $w_1\in\Span{1_N}$, $w_1 = c\cdot 1_N$ for some constant $c\in\RR\setminus\{0\}$.
Since the Gram-Schmidt process allows for $v_1^\prime \coloneqq \vert w_1\vert$, it follows that $v_1 = \vert c\vert \cdot 1_N / \Vert c\cdot 1_N\Vert = (\sqrt{N}/N)\cdot 1_N$.

Using the elements in $\beta$, let $\WV^\prime \coloneqq [v_1, v_2,...,v_N]\in\RR^{N\times N}$.
Similarly, let $\WD^\prime \coloneqq \diag{\lambda_1,\lambda_2,...,\lambda_N}\in\RR^{N\times N}$.
Recall $\lambda_1 = 0$ and $\lambda_p >0$ for every $p=2,3,...,N$.
Moreover, $(\WV^\prime)^{-1} = (\WV^\prime)^\top$ since the columns of $\WV^\prime$ form an orthonormal set.
Using the diagonalization of $\lap$, one has that\footnote{The Gram-Schmidt process and normalization do not affect the bases of the eigenspaces of $\lap$. 
Hence, $\{w_p,\lambda_p\}_{p\in\verts}$ and $\{v_p,\lambda_p\}_{p\in\verts}$ yield different diagonalizations of $\lap$.}
\begin{equation*}
    \lap = \WV^\prime \WD^\prime (\WV^\prime)^{-1} = \WV^\prime \WD^\prime (\WV^\prime)^\top = \sum_{p\in [N]} \lambda_p v_p v_p^\top = \lambda_1 v_1 v_1^\top + \sum_{p\in [N]\setminus\{1\}} \lambda_p v_p v_p^\top = \WV \WD \WV^\top,
\end{equation*}
such that $\WD=\diag{\lambda_2,\lambda_3,...,\lambda_N}\in\RR^{N-1\times N-1}$ is a positive definite matrix.
Recall $\tilde{\mathtt{e}}_k$ denotes the $k^\text{th}$ standard basis vector of $\RR^{N-1}$.
Since the columns of $\WV$ are pairwise orthogonal, one has that $\WV^\top v_p = \tilde{\mathtt{e}}_{p-1}$ for all $p = 2,3,...,N$.
Therefore,
\begin{equation*}
    \WV^\top \WV  = [\WV^\top v_2, \WV^\top v_3,..., \WV^\top v_N] = I_{N-1}.
\end{equation*}
Lastly, recall $\sproj = I_N - 1_N 1_N^\top / N$.
Note $\WV^\prime (\WV^\prime)^\top = I_N$ since $\WV^\prime$ is an orthonormal matrix.
It then follows that 
\begin{equation*}
\begin{aligned}
    \WV\WV^\top &= \sum_{p\in[N]\setminus\{1\}} v_p v_p^\top = \sum_{p\in[N]\setminus\{1\}} v_p v_p^\top + v_1 v_1^\top - v_1 v_1^\top = \sum_{p\in[N]} v_p v_p^\top - \frac{\sqrt{N}}{N}1_N \frac{\sqrt{N}}{N}1_N^\top \\
    &= \WV^\prime (\WV^\prime)^\top - \frac{1}{N}1_N 1_N^\top = \sproj.
\end{aligned}
\end{equation*}
Hence, the desired result follows.
\hfill\qedsymbol

\section{Proof of Lemma~\ref{lemma: Completeness and (t,j) Bounds}} \label{appen: completeness and (t,j) bounds}

\noindent \textbf{Item 1}: The proof is based on~\cite[Proposition 2.10]{Goebel.Sanfelice.ea2012}.
Consider the differential equation $\dot{\xi}=f(\xi)$ with initial condition $\phi(0,0)\in C\setminus D$.
\fred{Since the flow map $f$ in~\eqref{eqn: flow map} is globally Lipschitz}, there exists a nontrivial, maximal solution $\phi$ for $\hyb$ satisfying the initial condition.
Further, under the construction of $\hyb$, $G(D)\subset C\cup D$ implying that Item (c) in~\cite[Proposition 2.10]{Goebel.Sanfelice.ea2012} does not occur.
Since the single-valued flow map $f$ is Lipschitz continuous, Item (b) in~\cite[Proposition 2.10]{Goebel.Sanfelice.ea2012} does not occur.
Hence, $\phi$ is complete. 

\noindent \textbf{Item 2}: The following is guided by the proof of Lemma 3.5 in~\cite{Li.Phillips.ea2018}.
Fix a maximal solution $\phi$ of $\hyb$, and let $(t',j')\in \dom\phi$.

\textbf{Case I}: Suppose $(t,j)\in\dom\phi$ such that $0\leq t - t' \leq T_{\min}$ and $0 \leq j - j'$.
With respect to the dynamics in~\eqref{eqn: tau timer p} and~\eqref{eqn: rho timer r}, at most $N+M^\ast$ timers may jump twice during a flow interval of length $T_{\min}$.
This occurs when $\tau_p(t',j')=0$ with $\tau_p^+=T_{\min}$ for each $p\in\verts$ and $\rho_r(t',j')=0$ with $\rho_r^+=T_{\min}$ for each $r\in[M^\ast]$.\footnote{\label{foot: reset condition} Such a reset is possible when $T_1^p=T_3^r$ and $T_2^p=T_4^r$ for all $p\in\verts$ and $r\in[M^\ast]$.}
If a timer has a value in $(0,T_{\min}]$ at time $(t',j')$, then that timer jumps once during a flow interval of length $T_{\min}$.
Hence, the number of jumps between $(t',j')$ and $(t,j)$ is bounded as $j-j'\leq (N+M^\ast)(\tfrac{t-t'}{T_{\min}} + 1)$, which implies that
\begin{equation} \label{eqn: (t,j) lower bound}
    \left(\tfrac{j - j'}{N+M^\ast} - 1\right)T_{\min} \leq t - t'.
\end{equation}

\textbf{Case II}: Suppose $(t,j)\in\dom\phi$ such that $t - t' = T_{\max}$ and $0 \leq j - j'$.
With respect to the dynamics in~\eqref{eqn: tau timer p} and~\eqref{eqn: rho timer r}, at least $N+M^\ast$ timers must jump once during a flow interval of length $T_{\max}$.
This occurs when $\tau_p(t',j')=T_{\max}$ for each $p\in\verts$ and $\rho_r(t',j')=T_{\max}$ for each $r\in[M^\ast]$.\textsuperscript{\ref{foot: reset condition}}
If a timer has a value of $0$ at time $(t',j')$ and is reset to some value in $[T_{\min},T_{\max}]$, then that timer may jump at least twice during a flow interval of length $T_{\max}$.
Hence, the number of jumps between $(t',j')$ and $(t,j)$ is bounded as $ (N+M^\ast)\tfrac{t-t'}{T_{\max}} \leq j-j'$, which implies that
\begin{equation} \label{eqn: (t,j) upper bound}
      t - t' \leq \tfrac{j - j'}{N+M^\ast} T_{\max}.
\end{equation}
Combining~\eqref{eqn: (t,j) lower bound} and~\eqref{eqn: (t,j) upper bound} while setting $(t',j')=(0,0)$ yields the result in~\eqref{eqn: (t,j) inequality}.
\fred{Lastly, maximal solutions of $\hyb$ being complete and the inequality in~\eqref{eqn: (t,j) inequality} imply $\phi$ is non-Zeno.}

\hfill\qedsymbol

\bibliographystyle{elsarticle-num}
\bibliography{References}

\end{document}